\documentclass[article,pre,10pt,singlecolumn,longbibliography]{revtex4-1}
\usepackage{microtype}
\usepackage{graphicx}
\usepackage[colorlinks,linkcolor=blue,urlcolor=blue,citecolor=blue]{hyperref}

\usepackage{caption}
\usepackage{subcaption}
\usepackage{mathptmx}
\usepackage{times}
\usepackage{epsfig,amsopn}
\usepackage{graphicx}
\usepackage{bm,amsmath,amssymb}
\usepackage{natbib}
\usepackage{braket}
\usepackage{float}
\usepackage{wasysym}
\usepackage{enumerate}
\usepackage{hyperref}
\usepackage{comment}
\allowdisplaybreaks[1]
\usepackage[dvipsnames]{xcolor}

\def\be{\begin{equation}}
\def\ee{\end{equation}}
\def\bse{\begin{subequations}}
\def\ese{\end{subequations}}
\def\bcs{\begin{cases}}
\def\ecs{\end{cases}}
\def\bea{\begin{eqnarray}}
\def\eea{\end{eqnarray}}

\newcommand{\Aref}[1]{Appendix}%

\newcommand{\opunit}{\textrm{1}\kern-0.22em\textrm{l}}

\begin{document}
%\title{Equilibrium and non-equilibrium phase transitions in athermal random field Blume-Capel and Blume-Emery-Grifitths models}
\title{Glauber dynamics phase transitions in athermal random field Blume-Capel and Blume-Emery-Griffiths models}
%\title{Dynamical and equilibrium phase transitions in athermal random field Blume-Capel and Blume-Emery Grifitths models }
%\titel{Non-equilibrium transitions and hysteresis in athermal random field Blume-Capel and Blume Emery Grifitths models  }
\author{{\normalsize{}Sumedha$^{1, 2}$}
{\normalsize{}}}
\email{sumedha@niser.ac.in}

\author{{\normalsize{}Aldrin B E$^{1, 2}$}
{\normalsize{}}}
%\email{aldrin@niser.ac.in}

\affiliation{\noindent $^{1}$School of Physical Sciences, National Institute of Science Education and Research, Bhubaneswar, P.O. Jatni, 752050, India}

\affiliation{\noindent $^{2}$Homi Bhabha National Institute, Training School Complex, Anushakti Nagar 400094, }
\begin{abstract}
We solve the two models for Glauber dynamics and in equilibrium, both in the presence and absence of the external magnetic field on a complete graph. We compare the steady state of the Glauber dynamics with equilibrium and find that for low values of variance $R^2$ of the  Gaussian random field, the steady state of the Glauber dynamics depends on the initial state. Beyond a critical value $R_{c}$ the steady state does not depend on the initial state, and the equilibrium and non-equilibrium steady states coincide. The $R$ in random field models at zero temperature behaves similar to the temperature. Unlike the equilibrium, the location of both the continuous and first-order transitions can be obtained exactly for the Glauber dynamics steady state. While we consider only the ferromagnetic coupling between the spins, frustration is introduced  by considering repulsive biquadratic interaction for Blume-Emery-Griffiths model. For repulsive biquadratic interaction, we show that $R_c$ can become zero depending on the value of the crystal field. Interestingly, in some regime of the couplings, the model undergoes a crossover to a random field Ising model universality with $R_c=\sqrt{\frac{2}{\pi}}$. 

In the presence of uniform magnetic field, regions of first-order transition exhibit hysteresis under Glauber dynamics. These  models exhibit rectangular, hexagonal, parallelogram, wasp-waisted, and double hysteresis loops. We derive the shapes of hysteresis loops analytically by analysing the Glauber dynamics in the absence of disorder. We show that while the area under the hysteresis loop depends on $R$, the shape is determined by the behavior of the models at $R=0$. In particular, in the case of Blume-Emery-Griffiths model the hysteresis plots have regions of continuous and first-order transitions both, resulting in a rich phase diagram that depends non-trivially on the initial state. We also provide the equation for the value of the coercive field for the two models as a function of $R$.
\end{abstract}

\maketitle
%\tableofcontents

\section{Introduction}
\label{sec1}
The phase transitions play a central role in our understanding of a wide range of phenomena. While a framework exists to understand phase transitions in equilibrium, a similar universal framework for non-equilibrium steady states does not exist due to the lack of an underlying minimizing function in general. The non-equilibrium steady state depends crucially on the underlying dynamics. Glauber dynamics is the physical single  spin-flip dynamics used in many kinetic studies \cite{redner}.  It is a microscopic framework to model kinetics of many-body systems like magnets, social networks like opinion dynamics and consensus formation, as gradient descent in optimization and  neural networks and to model epidemics etc \cite{redner,social1,social2,gradient,epidemics}. While the finite temperature Glauber dynamics reaches the equilibrium steady state, at zero temperature the detailed balance is broken and the steady state of the dynamics is not an equilibrium state. A comparison of the zero temperature equilibrium state with the Glauber dynamics steady state at zero temperature hence gives us an important class of problems where we can compare the equilibrium and non-equilibrium phase transitions. The stochasticity in the model can be introduced via the introduction of quenched random field disorder. The  simplest model to consider is the random field Ising model (RFIM). The RFIM at $T=0$ has a order-disorder continuous non-equilibrium transition at  $R_c= \sqrt{2/\pi}$ on a complete graph. When an uniform external field is applied and varied quasi-statically, the model exhibit hysteresis for $R<R_c$\cite{sethna1,sethna}.

In this paper we focus on the three-state spin-$1$ models  in the presence of random field  disorder, namely the Random field Blume-Capel model (RFBCM) and the  Random field Blume-Emery-Griffiths model (RFBEGM). These models act as prototype models for many physical systems that cannot be modeled by two state systems like the Random Field Ising model (RFIM). The Blume-Capel model is the simplest model that exhibits a tricritical point. They have been used to  model systems such as superfluid mixtures, martensites, ferroelastics, metamagnets, multicomponent fluid mixture and so on~\cite{beg,metamag,ortin1,perez-reche,vasseur,fluid}.

Spin-models in the presence of external driving via a magnetic field also exhibit  hysteresis and are models of driven disordered systems. They are central to understanding a wide variety of physical phenomena, like crackling noise in magnetic materials, martensitic transformations, rigidity transitions in granular matter and earthquake dynamics along faults~\cite{crackling_noise, ava_martensites, rigidity,earthquakes}. The shape of the hysteresis loop is  connected to material properties \cite{dhar}. For example, double hysteresis loops are seen in ferroelectrics \cite{ferroelectricloops} and wasp-waisted hysteresis loops are seen in magnetic rocks \cite{wasp}. Recently, we also studied RFBEGM as a minimal and analytically tractable framework to study the  effect of frustration in driven disordered systems and showed evidence of frustration-induced blocking in non-abelian avalanche dynamics \cite{aldrin2}.

There are two order parameters in spin-$1$ models: the magnetisation (m) and the quadrupole moment (q). We obtain the  self consistent  equations for $m$ and  $q$ for the the Glauber dynamics steady state. We also obtain the zero temperature equilibrium free energy functional of the two models using large deviations using techniques developed in \cite{sumedhajana,soheli}. We extend the method to RFBEGM in this paper. We find that the self-consistent fixed point equations for the order parameters for equilibrium at zero temperature are the same as that of the Glauber dynamics. For Gaussian random fields  with mean $0$ and variance $R^2$,  the equilibrium phase diagram has a line of first-order transition at small $R$ which changes to a continuous transition at $R_{tcp}=\sqrt{\frac{2}{ e \pi}}=0.48394$ (with crystal field $\Delta=R$ at tricritical point(tcp)) \cite{soheli}, similar to the phase diagram of Blume-Capel model without disorder, with $R$ now  playing the role of temperature. We find the transitions in the Glauber dynamics steady state on the other hand depend on the initial state of the system and are path dependent. A comparison with the equilibrium phase diagram of the RFBCM shows that the Glauber dynamics steady state phase diagram coincides with the equilibrium phase diagram for values of $R$ larger than the value at the tricritical point. Below the tricritical point the Glauber steady state depends on the initial state. This is similar to the behavior of spin glasses as a function of temperature where it is known that below a critical temperature the distance between two initial states do not vanish in the steady state \cite{derrida}. The $R$ for athermal systems can be used as an appropriate parameter to tune noise in the system. We produce the RFBCM phase diagram for two initial states: $m=0$ and $m=1$. We also simulate the Glauber dynamics on a complete graph and show an excellent match between the simulations and the analytical phase diagram for the Glauber dynamics. Note that while for equilibrium the first-order transition point is determined numerically, for Glauber dynamics we obtain a closed form equation for the transition co-ordinates.

We then study the RFBEGM. Due to the presence of the biquadratic interaction the model has very rich behavior. RFBEGM with negative biquadratic coupling provides an excellent model to understand the interplay of frustration and disorder. For positive biquadratic coupling, its behavior is similar to RFBCM. We obtain the equation for the Glauber phase boundary and study it specifically for $m=0$ initial state. One of the striking results in the case of RFBEGM is the absence of phase transition in the presence of repulsive biquadratic interaction for positive crystal fields. We also discover that when the biquadratic coupling and the crystal field both are negative, the model exhibits a crossover in the Glauber dynamics steady state as a function of R. For small $R$, $m$ decays as if $R_c$ is zero. But once $q$ becomes sufficiently small it shows a phase transition at $R_c=\sqrt{2/\pi} \approx 0.8$ to the disordered state. We verify our analytical results with the simulations and find an excellent match.

In the presence of external magnetic field, the spin models undergo hysteresis under Glauber dynamics. The hysteresis loop in  the case of RFIM is  rectangular in shape. But the hysteresis loops seen in actual experiments take various shapes. We obtain the tally of all the possible hysteresis loop shapes for Blume-Capel and Blume-Emery-Griffiths models exactly. The models exhibit many different hysteresis shapes like double loop, wasp-waisted loops and hexagonal and parallelogram loops (which are similar to pot-bellied loops reported in literature). While these shapes have been seen in simulations and experiments\cite{ortin1,aldrin1,aldrin2}, our exact determination of the shapes allows us the understand the origin of these various hysteresis shapes. We show that while the area of the hysteresis loop depends on the value of $R$, the shape can be determined by solving the Glauber dynamics for $R=0$. We also derive the equation of transition in the presence of $H$ which gives the exact values of the coercive fields for RFBEGM and RFBCM for any $R$. In the presence of $H$, even the pure Blume-Emery-Griffiths model has a non-trivial phase diagram that depends on the initial state. We obtain it for two set of initial states: the one with all spins up and the other with all spins down. 

The paper is organised as follows: In Sec. \ref{sec2} we define the model. In Sec. \ref{sec3} we solve for the steady state of the Glauber dynamics. In Sec. \ref{sec4} we first obtain the free energy functional for arbitrary temperature for RFBEGM via large deviations and then take the $T \rightarrow 0$ limit. In Sec. \ref{sec5} we study both equilibrium and Glauber steady state phase diagrams for RFBCM in the absence of external field, the treatment is extended to RFBEGM in Sec. \ref{sec6}. In Sec. \ref{sec7} we study the two models in the presence of external field. In Sec. \ref{sec8} we conclude by summarising our results and discuss future directions.

\section{Model}
\label{sec2}
The Hamiltonian of the random field Blume-Emery-Griffiths model on a fully connected graph is given by:

\begin{eqnarray}
\mathcal{H}[\{s\}] &=& -\frac{J}{2N}\left( \sum_{i}s_i \right)^2 - \frac{K}{2N}\left( \sum_{i}s_i^2 \right)^2
- \sum_i (H + h_i)s_i + \Delta \sum_i s_i^2,
\label{hamiltonian}
\end{eqnarray}
Here $s_i$ are spin-$1$ variables that take values $\pm1,0$. $J$  is the ferromagnetic coupling that introduces co-operative interaction between $\pm 1$ states on two neighboring sites. $K$ is the bi-quadratic coupling term that couples the density of $\pm 1$ spins on the two neighboring sites. A positive value of $K$ introduces co-operative interaction by favoring $s_i^2 =1$ neighbouring states, while a negative value of $K$ disfavors them. The crystal field $\Delta$ is like the chemical potential controlling the density of $0$ spins and $H$ is the uniform external magnetic field. The $h_i$ are the quenched random fields at each site that are drawn from a continuous Gaussian probability distribution $p(h,R)=\sqrt{\frac{1}{2 \pi R^2}} e^{-h^2/2 R^2}$, with $R$ being the standard deviation of the distribution. Since we only consider the ferromagnetic interaction between the spins, we fix $J=1$ in the rest of the paper.

When the biquadratic coupling is absent, i.e $K=0$ the model is known as the Blume-Capel model. While for positive $K$, there are similarities in the behavior of the Blume-Capel and Blume-Emery-Griffiths models, for negative $K$ Blume-Emery-Griffiths model brings in the effect of frustration in the model. The two models especially differ for negative $K$ values. 
The models have two order parameters: the magnetisation $m=\langle s\rangle$  and the quadrupole moment $q=\langle s^2\rangle$.

Glauber dynamics is a single spin-flip dynamics. At zero temperature,  randomly chosen spin-flips if flipping lowers the energy of the system. The energy is given by the value of $\mathcal{H}[\{s\}]$. We solve for the order parameters $m$ and $q$ and obtain a set of coupled self-consistent equations in $m$ and $q$. We also study the Glauber steady state via simulations to verify analytical results. To solve for equilibrium, we obtain the generalised free energy functional of the model.

\section{Glauber dynamics}
\label{sec3}
 In Glauber dynamics spins update randomly based on the local influences of their neighbors. At zero temperature only the energy-lowering spin-flips are allowed in Glauber dynamics and the dynamics lacks detailed balance.  Hence the system relaxes to a non-equilibrium steady state. In the case of spin-$1$ models discussed here, the spin can flip to two possible states. The spin-flip that lowers the energy the most is chosen.

In order to study the Glauber dynamics of the RFBEGM we define two local field \cite{aldrin2}
\begin{equation}
L_1(s_k) = \frac{1}{N}\left(\sum_{i \neq k} s_i \right)+H+h_k
\label{lf1}
\end{equation}
\begin{equation}
	L_2(s_k) = \frac{1}{2N} + \frac{K}{2N} \left(2\sum_{i\neq k} s^2_i + 1 \right) - \Delta
	\label{lf2}
\end{equation}
The three possible energy change due to spin-flip under Glauber dynamics depending on the state of the spin are

1) The energy change caused due to the spin-flip at site $k$, $s_k \rightarrow -s_k$, with $s_k =  \pm 1$ 
\begin{equation}
 \delta E = 2 s_k L_1(s_k)
\end{equation}

2) The energy change caused due to the spin-flip at site $k$, $s_k \rightarrow 0$, with $s_k =  \pm 1$,
\begin{equation}
 \delta E = s_k L_1(s_k) + L_2(s_k)
\end{equation}

3) The energy change caused due to the spin-flip at site $k$, $0 \rightarrow s_k$, with $s_k =  \pm 1$,
\begin{equation}
\delta E = -s_k L_1(s_k) - L_2(s_k)
\end{equation}
We have recently shown that the dynamics depends on the sign of $L_2(s_k)$ for RFBEGM. The spin attains a value that depends only on the  state of its neighbors and not on its own present state. The rules of spin updates as obtained in \cite{aldrin2}, depending on values of $L_1(s_k)$ and $L_2(s_k)$ are as follows:
\begin{align}
	\text{If } L_2(s_k) < 0,\ \ L_1(s_k) \in
	\left\{
	\begin{array}{ll}
		\left( -\infty, L_2(s_k) \right) & \implies s_k = -1 \\[10pt]
		\left[ L_2(s_k), -L_2(s_k) \right) & \implies s_k = 0 \\[10pt]
		\left[ -L_2(s_k), +\infty \right) & \implies s_k = +1
	\end{array}
	\label{rule1}
	\right.
	\\[20pt]
	\text{If } L_2(s_k) \geq 0,\ \ L_1(s_k) \in
	\left\{
	\begin{array}{ll}
		(-\infty, 0) & \implies s_k = -1 \\[10pt]
		\left[0, \infty\right) & \implies s_k = +1
	\end{array}
	\right.
	\label{rule2} % Label the whole block as "rule"
\end{align}
In this paper we study the steady state by taking the  thermodynamic limit i.e, $N \rightarrow \infty$.  In the thermodynamic limit, $L_1(s _k)$ and $L_2(s_k)$ take the simpler form:
\begin{eqnarray}
    L_1(s_k) &=& m+ H+ h_k \\\nonumber
L_2(s_k) &=& K q -\Delta 
\end{eqnarray}
with  $m=\langle s\rangle=\sum_i s_i/N$ and $q =\langle s^2\rangle=\sum_i s_i^2/N$.

To study the steady state we calculate the probability for a randomly chosen spin to be in state $s$ and calculate the value of the order parameters $m$ and $q$. Let $P_s(K,\Delta,R,H)$ be the probability for a randomly chosen spin to be in state $s$. Then,
\begin{eqnarray}
    m &=& P_{+1}(K,\Delta,R,H)-P_{-1}(K,\Delta,R,H)\\
    q &=& P_{+1}(K,\Delta,R,H)+P_{-1}(K,\Delta,R,H)
    \label{mgd1}
\end{eqnarray}
The $P_s(K,\Delta,R,H)$ takes different forms depending  on the sign of $L_2(s)$. Hence we consider the  $L_2(s) \geq 0$ and $L_2(s) <0$  cases separately.
\subsection{{$L_2(s)\geq 0$}}
\label{sec31}

For $L_2(s) >0$, dynamics follows Eq. \ref{rule2} and the spin takes only two values: $\pm 1$. The probabilities are given by
\begin{align}
P_{+1}(K,\Delta,R,H) &= \int_{-m-H}^{\infty} d h\ p(h,R)\nonumber \\ &=\frac{1}{2} + \frac{1}{2} \text{erf} \left(\frac{m+H}{\sqrt{2} R}\right) 
\end{align}
and 
\begin{align}
P_{-1}(K,\Delta,R,H) &= \int_{-\infty}^{-m-H} d h\ p(h,R) \nonumber \\ &= \frac{1}{2} + \frac{1}{2} \text{erf} \left(\frac{-m-H}{\sqrt{2} R}\right) 
\end{align}
here $\text{erf}(x)= \frac{2}{\sqrt{\pi}}   \displaystyle\int_{0}^{x} e^{- t^2} dt $ is the error function.
As a result the average magnetisation per spin $m$ is 
\begin{align}
    m &= P_{+1}(K,\Delta,R,H)-P_{-1}(K,\Delta,R,H)\nonumber  \\
    &= \text{erf} \left(\frac{m+H}{\sqrt{2} R}\right)
    \label{mgd1}
\end{align}
Similarly the average value of the quadruple moment $q$ defined as the average value of $s^2$, is  
\begin{align}
    q &= P_{+1}(K,\Delta,R,H)+P_{-1}(K,\Delta,R,H)\nonumber \\
    &= 1
    \label{qgd1}
\end{align}
Since the Eq. \ref{mgd1} is the same as the steady state magnetisation equation  for RFIM under Glauber dynamics \cite{sethna1}, the model is exactly like the RFIM for this regime of couplings.

\subsection{{$L_2(s) < 0$}}
\label{sec32}

In this case dynamics follow Eq. \ref{rule1} and we get
\begin{align}
P_{+1}(K,\Delta,R,H) &= \int_{\Delta-K q-m-H}^{\infty} d h p(h,R)\nonumber \\ &= \frac{1}{2} + \frac{1}{2} \text{erf} \left(\frac{m+H-\Delta+Kq}{\sqrt{2} R}\right) 
\end{align}
and 
\begin{align}
P_{-1}(K,\Delta,R,H) &= \int_{-\infty}^{-\Delta+K q-m-H} d h p(h,R)\nonumber \\ &= \frac{1}{2} + \frac{1}{2} \text{erf} \left(\frac{-m-H-\Delta+Kq}{\sqrt{2} R}\right)
\end{align}
The value of $P_0 (K,\Delta,R,H) = 1- P_{+1}(K,\Delta,R,H)-P_{-1}(K,\Delta,R,H) $.

The steady state has $m$ and $q$ values given by
\begin{align}
   m &= \frac{1}{2} \text{erf} \left(\frac{x}{\sqrt{2} R}\right) + \frac{1}{2} \text{erf} \left(\frac{y}{\sqrt{2} R}\right)
   \label{mgd2}\\
   q & = 1+\frac{1}{2}\text{erf} \left(\frac{x}{\sqrt{2} R}\right)   -  \frac{1}{2}\text{erf} \left(\frac{y}{\sqrt{2} R}\right)
   \label{qgd2}
\end{align}
with $x=m+H-\Delta+Kq$ and $y=m+H+\Delta-Kq$.
These fixed point equations are obtained by averaging over all possible steady states for a fixed $L_2=(Kq-\Delta)<0$. Since $L_2$ can change when $q$ changes , the system can actually undergo a crossover to $L_2>0$ regime without changing $K$ and $\Delta$, We had recently shown that the dynamics for $K<\Delta$ violates no-passing property and can also become non-abelian \cite{aldrin2}. This perhaps causes this possible crossover in the steady state properties of the system.
\section{Equilibrium free energy: Derivation of the Rate function}
\label{sec4}
We now solve the model to get the equilibrium free energy. This  is done by using large  deviations approach  for performing the  quenched average following the method developed for random field spin models in \cite{sumedhajana}  and \cite{soheli}. We first derive the expression for finite temperature and then take the $T \rightarrow0$ limit.

Consider any random configuration $C_N$ of $N$ spins with $x_1 =\sum_{i=1}^N s_i/N$ and $x_2= \sum_{i=1}^N s_i^2/N$. The probability of occurrence of this configuration $P_{\mathcal{H},\beta}$ is proportional to $\exp(-\beta \mathcal{H})$, where $\beta=1/T$. The random variables $\left(\sum_{i=1}^N s_i,\sum_{i=1}^N s_i^2\right)$ satisfy the Large Deviation Principle (LDP) \cite{dembo,touchette,hollander} w.r.t to $P_{\mathcal{H},\beta}$. This implies that there exists a rate function $I(x_1,x_2)$ such that
\begin{equation}
P_{\mathcal{H},\beta}(C_N:x_1,x_2) \sim \exp(-N I(x_1,x_2))
\label{ldp}
\end{equation}
The rate function $I(x_1,x_2)$ is like the generalized free energy functional. Its minima gives the free energy of the system. 
The calculation of  $I(x_1,x_2)$  involves the following two main steps:
\begin{itemize}
\item Using the G{\"a}rtner-Ellis theorem \cite{dembo,touchette} and the law of large numbers, we first calculate the rate function $R(x_1,x_2)$ associated with the non-interacting part of the Hamiltonian denoted by $\mathcal{H}_{ni}$ and given by
\begin{equation}
\mathcal{H}_{ni} = - \sum_i (H + h_i)s_i + \Delta \sum_i s_i^2,
\label{hni}
\end{equation}
Then $R(x_1,x_2)$ is defined through
\begin{equation}
P_{\mathcal{H}_{ni},\beta}(C_N:x_1,x_2) \sim \exp(-N R(x_1,x_2))
\label{RHni}
\end{equation}
\item The probability $P_{\mathcal{H},\beta}(C_N:x_1,x_2)$ for the Hamiltonian is proportional to $\int_A e^{N f(x_1,x_2)} P_{\mathcal{H}_{ni},\beta}$, where $A$ is the subset of the all possible configurations, with a given $(x_1,x_2)$. The function $f(x_1,x_2)= \beta (x_1^2+K x_2^2)/2$ for Eq. \ref{hamiltonian}. The tilted large deviation principle \cite{hollander} then connects the two rate functions $I(x_1,x_2)$ and $R(x_1,x_2)$ through the relation
\begin{align}
I(x_1,x_2) &= R(x_1,x_2)-\frac{\beta x_1^2}{2} - \frac{\beta K x_2^2}{2}-\inf_{y_1,y_2} \left( R(y_1,y_2)-\frac{\beta y_1^2}{2} - \frac{\beta K y_2^2}{2}\right)
\label{rfI}
\end{align}
The probability measure $P_{\mathcal{H},\beta}(C_N:x_1,x_2)$ is the tilted version of $P_{\mathcal{H}_{ni},\beta}(C_N:x_1,x_2)$. 
\end{itemize}

Let us first calculate the rate function $R(x_1,x_2)$. Using the  G{\"a}rtner Ellis theorem it can be written as 
\begin{equation}
R(x_1,x_2) =  \sup_{y_1,y_2}  \{x_1 y_1 +x_2 y_2 -\Lambda(y_1,y_2)\}
\label{rfR}
\end{equation}
provided that the scaled cumulant generating function $\Lambda(y_1,y_2)= \lim_{N \rightarrow \infty} \Lambda_N(y_1,y_2)/N$, 
is differentiable \cite{dembo,touchette}.

The function $\Lambda_N(y_1,y_2)$ is the log cumulant generating function for the probability distribution $P_{\mathcal{H}_{ni},\beta}$. In the thermodynamic limit we get
\begin{equation}
\Lambda_N(y_1,y_2) = \log \left\langle \exp\left( y_1 \sum_{i=1}^N s_i + y_2 \sum_{i=1}^N s_i^2\right) \right\rangle_Q
\end{equation}
Here $\langle ... \rangle_Q$ represents the expectation value w.r.t. the probability distribution $Q \propto e^{-\beta \mathcal{H}_{ni}}$, which is a product measure over the probability distributions $Q_i$ for the non-interacting spins. This yields after summing over $i$ and using the law of large numbers, $\Lambda(y_1,y_2)$ as 
\begin{align}
\Lambda(y_1,y_2) &= \left\langle  \log \left[1+2 e^{(y_2-\beta \Delta)} \cosh{(y_1+\beta H+\beta h)} \right] \right\rangle_h
\label{lambdainf}
\end{align}
where $\langle... \rangle_h$ represents the average over the quenched field distribution, $p(h,R)$.

The $(y_1^*,y_2^*)$ extremise the r.h.s of Eq. \ref{rfR}. Both $y_1^*$ and $y_2^*$ are functions of $x_1$ and $x_2$ and  are given by the solutions of the equations:
\begin{align}
x_{1,2} &= \frac{\partial \Lambda(y_1,y_2)}{\partial y_{1,2}}
\end{align}
The rate function $R(x_1,x_2)$ can then be written as
\begin{align}
R(x_1,x_2) = g(x_1,x_2) - \inf_{x_1,x_2} g(x_1,x_2)
\end{align}
where
\begin{equation}
g(x_1,x_2) = x_1 y_1^* +x_2 y_2^* -\Lambda(y_1^*,y_2^*)-\frac{\beta (x_1^2+Kx_2^2)}{2}
\end{equation}
we don't have to worry about the infimum as  that is just a constant and plays no role in determining the extremas.
The full rate function is then given by substituting $R(x_1,x_2)$ in Eq. \ref{rfI}. In the thermodynamic limit, the probability 
$P_{\mathcal{H},\beta}(C_N:x_1,x_2)$  in  Eq. \ref{ldp} is dominated by the minimum of $I(x_1,x_2)$. The equations $\frac{\partial I}{\partial x_1}=0$ and $\frac{\partial I}{\partial x_2}=0$ yields $y_1^* = \beta x_1$ and $y_2^*=\beta K x_2$. The rate function is like a generalized 
free energy functional in that its minimum $\frac{1}{\beta} \inf_{x_1,x_2} I(x_1,x_2)$ provides the free energy of the system. 
By susbtituting $y_1^*$ and $y_2^*$ in Eq. \ref{rfI} we get the rate function as
\begin{align}
I(m,q) &=\frac{\beta}{2} (m^2+K q^2)- \left\langle  \log \left[1+2 e^{\beta (K q-\Delta)} \cosh{\beta (m+H+h} \right] \right\rangle_h
\label{qrf}
\end{align}
Since $x_1$ and $x_2$ are now the average values of $s$ and $s^2$ after quenched averaging, we have replaced $x_1$ and $x_2$ by $m$ and $q$ respectively in the above expression.

Equation \ref{qrf} is the general expression of the free energy functional for the RFBEGM model on a fully connected graph for an arbitrary distribution of disorder.  The free energy of the system is equal to $\frac{1}{\beta} \inf_{m,q} I(m,q)$. Here $m$ and $q$ are the magnitudes of magnetisation and quadrupole moment respectively and are the two order parameters of the system. They are given by
\begin{equation}
m=\left\langle \frac{2 e^{\beta (K q-\Delta)} \sinh \beta (m+H+h)}{1+2 e^{\beta (K q-\Delta)} \cosh \beta (m+H+h)} \right\rangle_h
\label{mfp}
\end{equation}
and 
\begin{equation}
q=\left\langle \frac{2 e^{\beta (K q-\Delta)} \cosh \beta (m+H+h)}{1+2 e^{\beta (K q-\Delta)} \cosh \beta (m+H+h)} \right\rangle_h
\label{qfp}
\end{equation}

\subsection{Zero Temperature RFBEGM}
\label{sec41}
In this paper we focus on the $T=0$ phase diagram. We now derive the ground state rate function defined as $f(m,q) = \lim_{\beta \rightarrow \infty} \frac{1}{\beta} I(m,q)$. The expression of $f(m,q)$ depends on $z=K q-\Delta$ and is different for $z \leq  0$ and for $z > 0$.

\subsubsection{$z>0$}

For $z>0$, from  Eq. \ref {qfp} , we see that $q \rightarrow 1$ in the limit of $\beta \rightarrow \infty$. 

Substituting $q=1$ in Eq. \ref{qrf} and taking the disorder average for the Gaussian quenched fields, we get 
\begin{align}
f(m) &= \frac{1}{2} (m^2+\Delta-K)-(m+H) \text{erf}\left(\frac{m+H}{\sqrt{2} R}\right) - R\sqrt{\frac{2}{\pi}} \text{exp}\left(-\frac{(m+H)^2}{2 R^2}\right)
\label{zp0T}
\end{align}

The fixed point equation for $m$ is
\begin{equation}
m =\text{erf} \left(\frac{m+H}{\sqrt{2}R} \right)
\label{mfp1}
\end{equation}
The $\inf_{m} f(m)$ is the free energy of  the system. Note that  a solution of RFIM  at $T=0$ yields the same equations as  Eqs.  \ref{zp0T} and \ref{mfp1} with $\Delta=0$. The condition $z>0$ is always  satisfied for $K  > \Delta$.  Hence  for $z>0$, the coupling $K$ and crystal field $\Delta$ play  no role.  Both equilibrium and non-equilibrium steady states are the same as that of RFIM.

\subsubsection{$z \leq 0$}
For $z <0$, $f(m,q) = \lim_{\beta \rightarrow \infty} \frac{1}{\beta} I(m,q)$ comes out to be

\begin{align}
f(m,q) = \frac{m^2}{2}+\frac{K q^2}{2}-\frac{x}{2} \text{erf}\left(\frac{x}{\sqrt{2} R}\right)-\frac{y}{2} \text{erf}\left(\frac{y}{\sqrt{2} R}\right) \nonumber \\
- z-\frac{R}{\sqrt{2 \pi}} \text{exp}\left(-\frac{x^2}{2 R^2}\right)-\frac{R}{\sqrt{2 \pi}} \text{exp}\left(-\frac{y^2}{2 R^2}\right)
\label{zn0T}
\end{align}
where $z=Kq-\Delta$, $y=-Kq+\Delta+m+H$ and $x=K q -\Delta + m+H$.
 
The fixed points of $f(m,q)$ for a given $(K,\Delta,R)$ are given by the solutions of $\partial f(m,q)/\partial m =0$ and $\partial f(m,q)/\partial q =0$. These give the fixed point equations to be 
\begin{align}
m =\frac{1}{2}  \text{erf} \left(\frac{x}{\sqrt{2}R} \right)+\frac{1}{2}  \text{erf} \left(\frac{y}{\sqrt{2}R} \right)
\label{znm}
\\
q = 1 + \frac{1}{2} \text{erf} \left(\frac{x}{\sqrt{2} R}\right)-\frac{1}{2} \text{erf} \left(\frac{y}{\sqrt{2} R}\right)
\label{znq}
\end{align}

\section{Random field Blume-Capel Model}
\label{sec5}
For $K=0$, the rate function does not depend on $q$ and is  a function only of $m$. This is the Blume-Capel model. We study the zero temperature phase transitions in equilibrium and in non-equilibrium steady state of the Glauber dynamics in this section in the absence  of external field $H$. Hence in this section we have fixed $K=0$ and $H=0$ in Eq. 1.

\subsection{Equilibrium phase diagram}
\label{sec51}
In the absence of disorder ($R=0$), the energy per spin is:
\begin{equation}
E =-\frac{m^2}{2} +\Delta q
\end{equation}
There is hence a first-order equilibrium transition at $\Delta=\frac{1}{2}$, where the $(|m|,q)$ changes from $(1,1)$ to $(0,0)$ via a first-order transition. 

For non zero $R$ with $\Delta \leq  0$  rate function is  given by Eq. \ref{zp0T} with $K=0$. All spins take only two values: $\pm 1$. Since the minima of the rate function does not depend $\Delta$  in this case, the system behaves like RFIM. The Eq. \ref{mfp1}  with $H=0$ has multiple solutions for $R<R_c = \sqrt{2/\pi}$. For $R \geq R_c$, $m=0$ is the only solution. The model undergoes a continuous transition at $R_c$ similar to the finite temperature ferromagnetic-paramagnetic transition in the Curie Weiss model. The global minima of the rate function changes from $m=0$ to $m \neq 0$ at $R=R_c$. Since the Glauber dynamics equation for steady state value of $m$, Eq. \ref{mfp1} is the same, the Glauber dynamics steady state is the same as the equilibrium state of the system. Note that this is true only when external  magnetic  field $H=0$. For  $H  \neq 0$,  the  Glauber  dynamics steady state differs from the  equilibrium state even  for $\Delta \leq 0$. This is discussed in Sec. \ref{sec7}

For $\Delta > 0$, all three spin states:  $\pm 1,  0$ are relevant. This changes the behaviour of the  model non-trivially. In the presence of quenched random fields with Gaussian distribution, the $T=0$ rate function for RFBC for $\Delta >0$  via Eq. \ref{zn0T}  is

\begin{align}
f(m,q) = \frac{m^2}{2}+\Delta-\frac{x}{2} \text{erf}\left(\frac{x}{\sqrt{2} R}\right)-\frac{y}{2} \text{erf}\left(\frac{y}{\sqrt{2} R}\right) \nonumber \\ -\frac{R}{\sqrt{2} \pi}\text{exp}\left(-\frac{x^2}{2 R^2}\right)-\frac{R}{\sqrt{2 \pi}} \text{exp}\left(-\frac{y^2}{2 R^2}\right)
\label{gaussgroundfree}
\end{align}
here $x=-\Delta-m$ and $y=-\Delta + m$.
%\begin{eqnarray}\label{gaussgroundfree}
%f(m) & =  & \frac{ m^2}{2}  - \frac{m}{2} \Bigg (  \text{erf}{ \Big (\frac{m + \Delta}{\sqrt{2} R}} \Big )  -  \text{erf}{ \Big (\frac{-%m + \Delta}{\sqrt{2} R}} \Big )   \Bigg ) 
	%		+  \frac{\Delta}{2} \Bigg ( 2 -  \text{erf}{ \Big (\frac{-m + \Delta}{\sqrt{2} R}} \Big )  -  \text{erf}{ \Big (\frac{m + \Delta}{\sqrt{2} R}} \Big ) \Bigg ) \nonumber \\
		%  &	- & \frac{R}{ \sqrt{ 2 \pi}} \Bigg (  \exp \left({- \frac{(-m + \Delta)^2}{2 R^2}} \right) +  \text{exp} \left({- 5\frac{(m + \Delta)^2}{2 R^2}}\right) \Bigg ) 
         % \label{bcI}
	%\end{eqnarray}
The $m$ and $q$ values given by the fixed point equations:

\begin{align}\label{magn}
m = \frac{1}{2} \Bigg (   \text{erf}{ \Big (\frac{m + \Delta}{\sqrt{2} R}} \Big )  -  \text{erf}{ \Big (\frac{-m + \Delta}{\sqrt{2} R}} \Big )   \Bigg )
\end{align}

\begin{align}\label{dens}
q = 1-\frac{1}{2} \Bigg (  \text{erf}{ \Big (\frac{m + \Delta}{\sqrt{2} R}} \Big )  + \text{erf}{ \Big (\frac{-m + \Delta}{\sqrt{2} R}} \Big )   \Bigg )
\end{align}

For  the case of Gaussian distribution of the random fields, the equilibrium phase diagram was obtained in \cite{soheli}, which we reproduce below. The phase diagram in the $\Delta-R$  plane is similar to $\Delta-T$  phase diagram of the pure Blume-Capel  model: There  is a line of  second-order  transitions, which meets a  line of first-order transition  at $R=R_{tcp}$. To obtain the phase diagram, expand $f(m,q)$ in  Eq. \ref{gaussgroundfree} around $m=0$. The expansion till sixth order in $m$ is
\begin{equation}
    f(m,q) =a_0+a_2 m^2+a_4  m^4+a_6 m^6
\end{equation}
with
\begin{align}
a_2 &= \frac{1}{2}-\frac{e^{\frac{-\Delta^2}{2 R^2}}}{\sqrt{2 \pi} R}\nonumber \\
a_4 &= - e^{\frac{-\Delta^2}{2 R^2}}\frac{(\Delta^2 -R^2)}{ 12 \sqrt{2 \pi} R^5} \nonumber \\
a_6 & =  - e^{\frac{-\Delta^2}{2 R^2}}\frac{(\Delta^4 -6 \Delta^2 R^2 + 3 R^4)}{ 360 \sqrt{2 \pi} R^9} 
\end{align}
    %    \begin{eqnarray}
		%a_8^0  & =& - e^{\frac{-\Delta^2}{2 R^2}}\frac{(\Delta^6 - 15 \Delta^4 R^2 + 45 \Delta^2 R^4 - 15 R^6)}{ 20160 \sqrt{2 \pi} s^{13}} 	
	%\end{eqnarray}
	
%\begin{figure}
%\centering
%\includegraphics[width=0.55\textwidth]{phasetransitiongaussian.eps}       
%\caption{Ground state phase diagram for the Gaussian field distribution. Dotted line is the line of first-order transitions and %solid line is the line of second-order transitions. Solid circle is the TCP. There is one ordered phase ($m \neq 0$) and one disordered phase ($m=0$) in the phase diagram. The transition is first-order for small $R$. As $R$ increases, the transition changes to second-order at a TCP with the coordinates $R_{TCP}=\Delta_{TCP}=\sqrt{\frac{2}{e \pi}}.$ }  
%   \label{fig3}
%\end{figure}
%There is one ordered phase with $m \neq 0$ and one disordered phase with $m=0$. The quadrupole moment $q$ changes continuously from $q=1$ to $q=0$ as $\Delta $ goes from $0$ to $\infty$. Thus there is no transition in $q$.
The continuous transition occurs when $a_2=0$, provided $a_4>0$. Since $a_4$  becomes $0$ when $\Delta^2=R^2$, the model has a line of continuous transitions for $R>\Delta$ given by the equation
\begin{equation}
R_c = \sqrt{\frac{2}{\pi}} \exp \left({\frac{-\Delta_c^2}{2 R_c^2}}\right)
\end{equation}
The line ends at $\Delta_{tcp}=R_{tcp}=\sqrt{\frac{2}{e \pi}} =  0.483941$. Since $a_6>0$ at  this point, the Taylor expansion is valid and this is a tricritical point where the line of second-order transitions terminates.

For $R<R_{tcp}$, the model has a first-order transition and we have to study the entire $f(m)$. The values of $\Delta,R$ where the global minima of the the function $f(m)$ shifts from $m=0$ to $m  \neq 0$ is the point of equilibrium first-order transition. The resultant phase diagram is shown in Fig. \ref{fig2}.

\subsection{Non-equilibrium phase diagram}
\label{sec52}

%The Glauber dynamics steady state for $R=0$ and $\Delta  \leq 0$, is the same as the equilibrium state of the system since $s_i$ takes only $\pm 1 $ values.  On the other hand the  steady state depends  on the initial  state for $\Delta >0$. For example, if we start with $m=0$ then  the steady state is $m=0$ $\forall \Delta  \geq  0$.  On the other  hand starting with  $|m|=1$ results in a transition from $|m|=1$ steady  state to $m=0$  steady state  at $\Delta=1$.  Other starting  values of $|m|$ give different transition points. This is illustrated in Fig 1 (a), where we have simulated the Glauber dynamics at $R=0$ for two different initial states. We can see that the trajectory depends on the initial state of the system.

The system under Glauber dynamics is updated according to the values of the local fields $L_1$ and $L_2$ defined in Eqs. \ref{lf1} and \ref{lf2}. For Blume-Capel model, since $K=0$, $L_2(s_k) = -\Delta$ $\forall k$. Hence for $\Delta<0$ the dynamical update rules under Glauber dynamics follow Eq. \ref{rule2} . The crystal field $\Delta$ does not play a role, and the steady state has the same properties as equilibrium RFIM: At $R = R_c=\sqrt\frac{2}{\pi}$, there is a continuous transition in $m$ from ferromagnetic to paramagnetic state. The Glauber dynamics steady state behaves similar to the equilbrium  state  for $\Delta \leq 0$.

We now focus on $\Delta>0$ in this section. First let us consider the case without disorder ($R=0$). The steady state under Glauber evolution now  depends on the initial state. For example, if we start with a state with $s_i=0 ~\forall i$ at large positive $\Delta$, the condition $L_2(i)<L_1(i)<-L_2(i)$ is always satisfied, and hence via Eq. \ref{rule1} , there is no change and the system continues to be in the state $m=0$ for all values of $\Delta$. 

On the other hand, if we start with $s_i=-1~\forall i$, then initially $L_1(i) =-1+\frac{1}{N}$ and $L_2(i)= -\Delta+\frac{1}{2N}$ $\forall i$. But at $\Delta_c=1+\frac{1}{N}$, $L_1(i)=L_2(i) \forall i$ and the first spin-flips to $0$. This starts an avalanche  where all the spins flip to $0$ at $\Delta_c$ before the system reaches the steady state. The two cases are shown in Fig. \ref{fig2}(a). The plots are generated by simulating Glauber dynamics quasi-statically by increasing $\Delta$ from $0$ in small increments. After  every increment in $\Delta$ the system is evolved till it reaches a steady state. 

We can hence conclude that the steady state is a  function of the initial state for $\Delta > 0$  under Glauber evolution. The transition point for the two initial states are different from each other and also from  the equilibrium value  of $\Delta=1/2$ discussed in the previous section.

\begin{figure*}[htbp]
\centering
\includegraphics[width=0.6\textwidth]{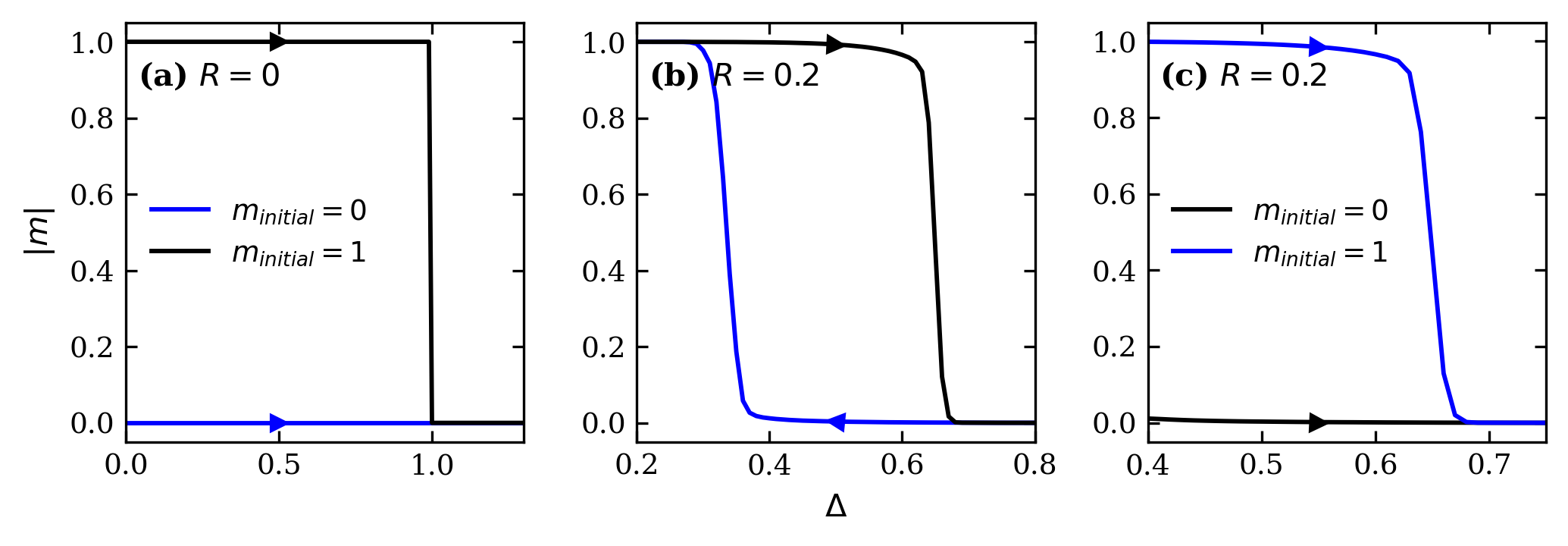}
\caption{Plot of steady state magnetisation($|m|$) as a function of $\Delta$, with  $\Delta$ increased quasi-statically. a) No disorder case with $R=0$ for two initial states:$m=0$ and  $|m|=1$ with increasing $\Delta$  from $\Delta=0$. For $m=0$ initial state the system continues to stay in $m=0$ state for all values of  $\Delta$, whereas for $|m|=1$ initial state there is a first-order transition to $m=0$ at $\Delta=1$; b) For $R=0.2$  system is evolved from $|m|=1$ initial state at $\Delta=0$. This state transitions to $m=0$ state at $\Delta \approx 0.64$. On the other hand the state with $m=0$ at  $\Delta=1$  on decreasing $\Delta$ transitions to $m=1$ at $\Delta \approx 0.33$ ; c) For $R=0.2$ and starting at $\Delta=0.4$ while the $m=0$ initial state continues to be $m=0$ with increasing $\Delta$; the $|m|=1$ initial state transitions to $m=0$ state around $\Delta \approx 0.64$. All the simulations are performed on a complete graph with $N=1000$ spins and averaged over $1000$ disorder realizations. }
\label{fig1}
\end{figure*}

\begin{figure*}[htbp]
\centering
\includegraphics[width=0.5\hsize]{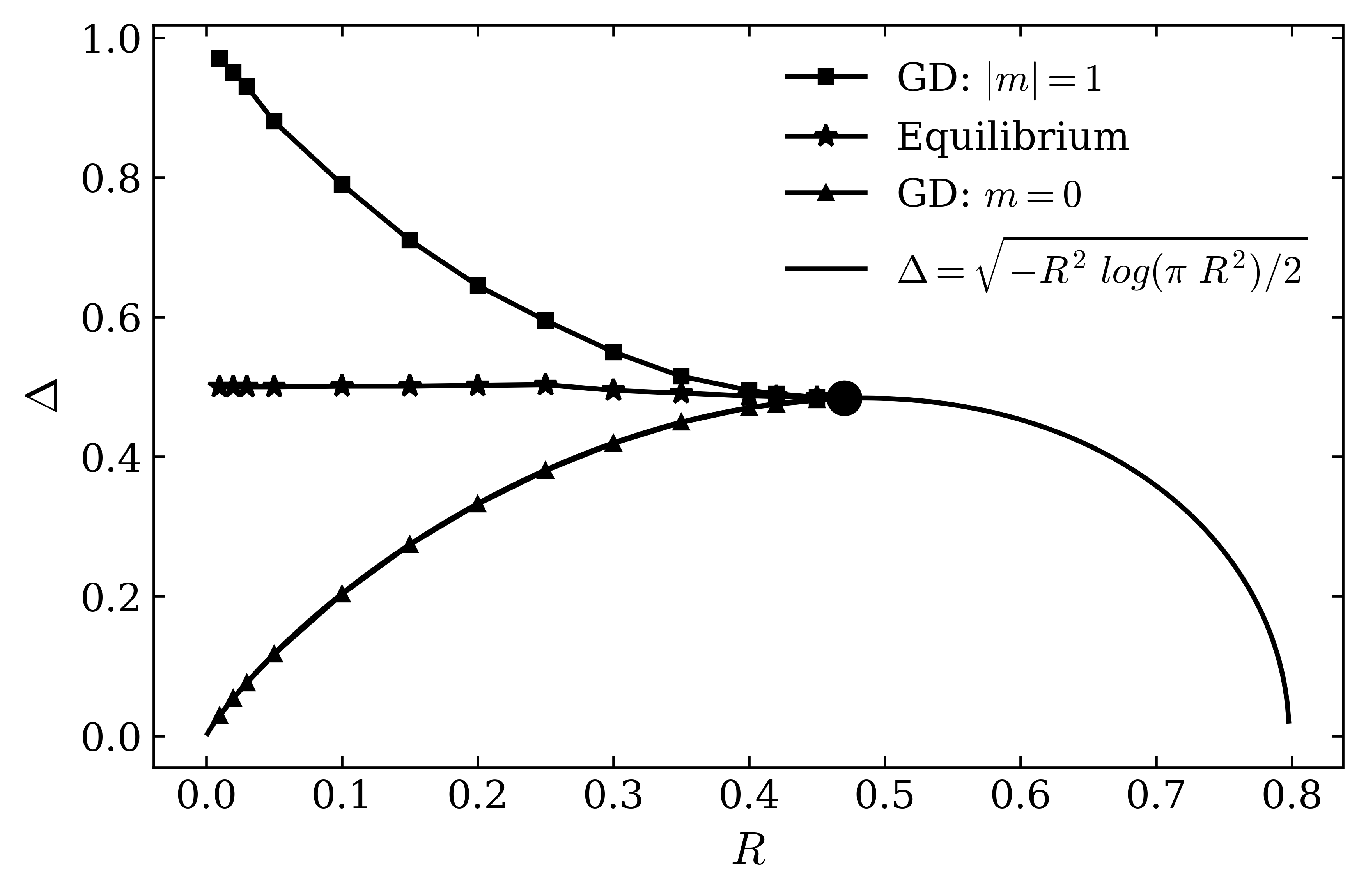}
\caption{Phase diagram of RFBCM under Glauber dynamics with initial state  $|m|=1$ and $m=0$ and the equilibrium phase diagram are shown. The black solid circle represents the tricritical point with $R_{tcp}=\Delta_{tcp}=0.483$. For $R \geq R_{tcp}$, the three phase  diagrams  coincide and there is a continuous transition  between $m=0$ and $|m|=1$ states. On the other hand for $R<R_{tcp}$, even though the nature of the transition ( first-order )  is the same in all the cases, the point of transition are different.}
\label{fig2}
\end{figure*}
\begin{figure*}[htbp]
\centering
\includegraphics[width=0.6\hsize]{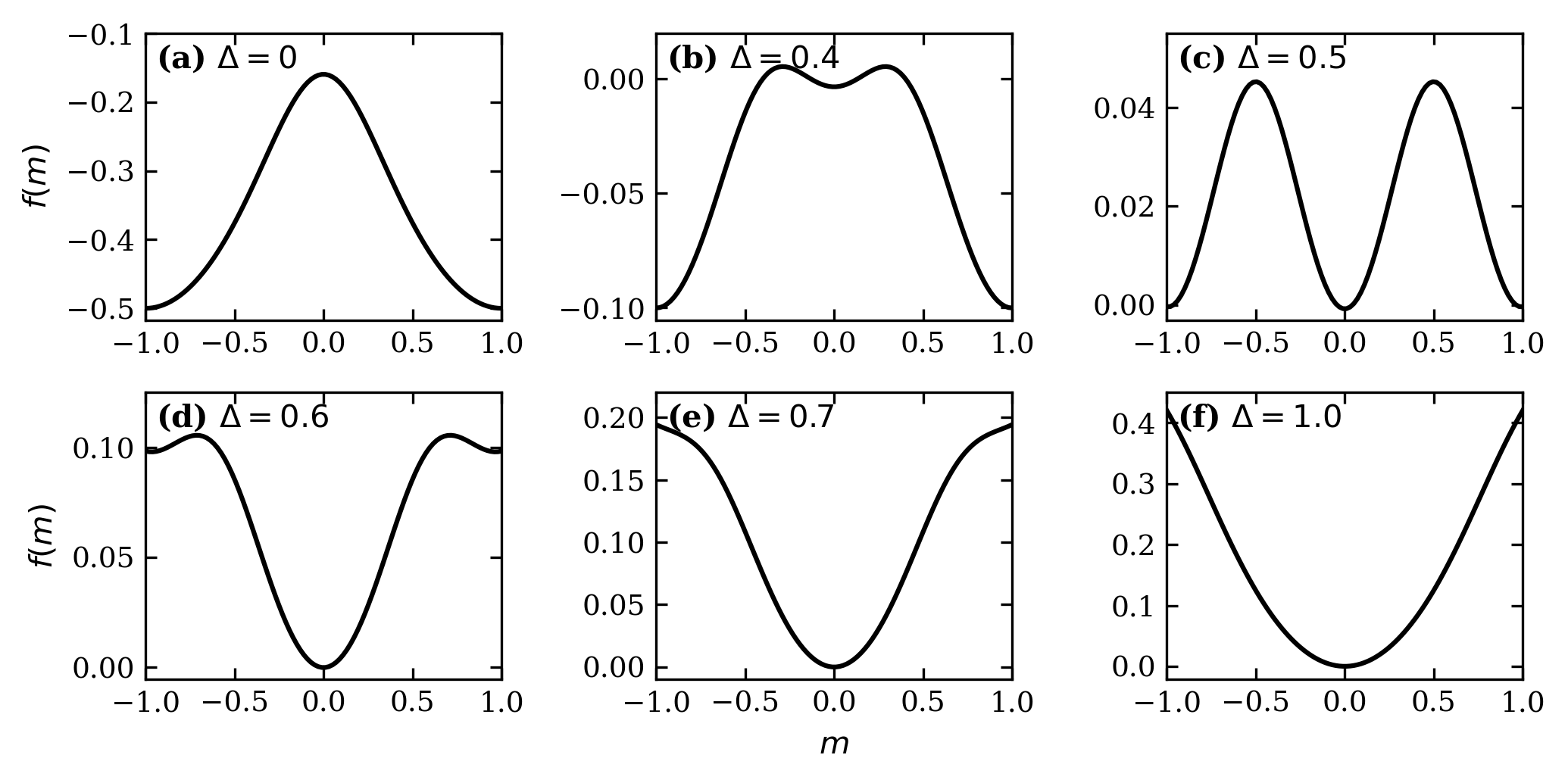}
\caption{For $R=0.2$, the function $f(m)$ is plotted for few representative values of $\Delta$.}. 
\label{fig3}
\end{figure*}
%\begin{figure*}[htbp]
%\centering
%\includegraphics[width=0.45\hsize]{fig_2.pdf}
%\caption{abc}
%\label{fig2}
%\end{figure*}
For $\Delta > 0$, under Glauber evolution even though we do not have a minimising function like $f(m)$, interestingly the fixed point equations for  equilibrium and steady-state fixed point equation for Glauber dynamics are the same. Since Glauber dynamics  only allows energy-lowering moves at $T=0$, the  system prepared in a local minima basin of a fixed point of the equilibrium free energy functional is at that fixed point in the steady state as long as the fixed point continues to be  a point of local minima. Since  the state  of  the system is always given by the global minimum if the equilibrium  is achieved, this results in the possibility of a different steady state under Glauber dynamics than the equilibrium states.

Hence the transition under  Glauber dynamics occurs when a local minima of $f(m)$ becomes  a maxima. Since for $R \geq R_{tcp}$ the global and local minima are the same, the Glauber steady state is the equilibrium state. In this case there is no distinction between equilibrium and non-equilibrium phase transition. But for $R<R_{tcp}$, there are multiple minimas and a state prepared in the basin of a particular minima will continue in that state till the fixed point stops being a point of minima of $f(m)$. For example, if we prepare the system in the state $m=0$ at large $\Delta$ and reduce $\Delta$ quasi-statically, the $m$ continues to be $0$ much below the critical equilibrium value of $\Delta$. The condition of transition near a fixed point under Glauber dynamics depends only on the local stability of the fixed point for $m$. 

The fixed point equation for $m$ for Glauber dynamics is
\begin{equation}
m = \frac{1}{2} \text{erf} \left(\frac{m-\Delta}{\sqrt{2} R}\right) +\frac{1}{2} \text{erf} \left(\frac{m+\Delta}{\sqrt{2} R}\right)
\label{gfp}
\end{equation}
This is also the equation obtained by equating $\partial f(m)/\partial m=0$. Because of this,  in the case of Glauber dynamics, even though we do not know the underlying equivalent of the function $f(m)$, but the second derivative of that function with respect to $m$ is the same as $\partial^2 f(m)/\partial m^2$ at the fixed  points. Defining $g(m,\Delta,R)=\partial^2 f(m)/\partial m^2$, we get
\begin{equation}
    g(m,\Delta,R) = 1-\sqrt{\frac{1}{2 \pi R^2}} \left(e^{-(m-\Delta)^2/2 R^2}+e^{-(m+\Delta)^2/2 R^2}\right)
\end{equation}
Hence a fixed point is a local minima if  $g(m,\Delta,R)>0$ and is locally a maxima if $g(m,\Delta,R)<0$. Hence under Glauber dynamics, $g(m_{fp},\Delta,R)=0$ gives the phase transition point for a system prepared in a initial state which is a fixed point of Eq. \ref{gfp}. This is true for continuous as well as first-order transitions under Glauber dynamics.  For an arbitrary initial state, the transition is determined by the $m_{fp}$ of the basin in which the initial state falls.

For a system prepared in the state $m=0$ or with $m$ which falls in the basin of $m=0$, there is a transition in the steady state to a non zero $m$ at $\Delta$ and $R$ that satisfy the condition:
\begin{equation}
    g(0,\Delta,R) = 1-\sqrt{\frac{2}{\pi R^2}} \text{exp}^{-\Delta^2/2 R^2}=0
    \label{bcmzerofp}
\end{equation}

Since this equation is same as the equation for critical point under equilbirum, the phase diagram between $m=0$ state and $m \neq 0$ state for a state prepared in $m=0$ state will be same as the equilibrium phase diagram for $R \geq R_{tcp}$. For  $R<R_{tcp}$, while the equilibrium state is obtained by finding the global minimum of $f(m)$ numerically, the first-order transition under Glauber dynamics is given by the Eq. \ref{bcmzerofp}. The phase boundary for $m=0$ initial state under Glauber dynamics is shown in Fig. \ref{fig2} and is given by 
\begin{equation}
    \Delta_c = \sqrt{-R_c^2 \text{log}(\pi R_c^2/2)}
\end{equation}
Similarly for a system prepared in $m=1$ state the phase boundary is different for $R<R_{tcp}$, as shown in Fig. \ref{fig2}.

As a consequence, for $R<R_{tcp}$, the transition is  path dependent. We illustrate this in Fig. \ref{fig1}(b)  and \ref{fig1}(c) for $R=0.2$. The system prepared in the state $m=1$  at $\Delta=0$ transition  to  $m=0$  state at around $\Delta \approx 0.64$  consistent with the $m=1$ Glauber dynamics phase boundary in Fig.  \ref{fig2}. On the other hand the system prepared in $m=0$  state at large positive $\Delta$ transitions to $|m| \neq 0$ state around  $\Delta=0.33$, which is consistent with the $m=0$ Glauber  dynamics phase  boundary in Fig. \ref{fig2}. Similarly, in Fig. \ref{fig1}(c),  we studied  two initial states $m=0$ and $m=1$ prepared at $\Delta=0.4$. While $m=0$ does  not change,  the $m=1$ state transitions  to $m=0$ only  at $\Delta \approx 0.64$. In  Fig. \ref{fig3} we have plotted $f(m)$ for different values of $\Delta$ for $R=0.2$. The $m=0$ local minima first appears around $\Delta =0.33$ and becomes a global minima at $\Delta=0.5$. Similarly, local minima around $m=1$ becomes unstable  around $\Delta=0.64$. As  a result, on increasing $\Delta$  quasi-statically from $\Delta=0$ for  any initial state, results in a  transition at $\Delta=0.33$, which is the value given by Eq.  \ref{bcmzerofp}. On the  other hand, starting with large positive $\Delta$, with any initial state the transition to $m  \neq 0$  state occurs at $\Delta_c= 0.64$, given by the phase boundary for $m=1$ state in Fig. \ref{fig2}. 
%In general, a state prepared between $0.2<\Delta<0.64$ will transition at different points in the phase space, that depend on the initial state value of $m$ and $\Delta$. %This is also illustrated in Fig.1 (b) and (c), where the $m$ is obtained by simulating Glauber dynamics on a fully connected graph with $N=1000$ spins.

\section{Random Field Blume-Emery-Griffiths Model (RFBEGM)}
\label{sec6}
We saw in the above section that the Glauber dynamics phase diagram and the equilibrium phase diagram differ only in the region of first-order transition. In this region for the equilbirium phase diagram, while the first-order transition had to  be obtained  numerically, the criterion of the existence of the local minimum is sufficient to get the Glauber phase boundary.

For $K > \Delta$, the $z=K q-\Delta>0$ as shown in Sec. \ref{sec41}, the model is equivalent to the RFIM with $q=1$. 

For  $K  \leq \Delta$, $z= K q-\Delta \leq 0$, and the free energy is given by the minima of the function $f(m,q)$ in Eq. \ref{zn0T} with fixed point equations for $m$  and $q$  being defined via Eq. \ref{znm} and \ref{znq}. Since $f(m,q)$ is now a function of two variables, the conditions  for the stability of a fixed point are now given  by the condition on the determinant of the  Hessian matrix (L). The Hessian matrix is
\begin{align}
L
&=
    \begin{bmatrix}
      \partial^2 f(m)/\partial m^2 & \partial^2 f(m)/\partial m \partial q \\
        \partial^2 f(m)/\partial q\partial m & \partial^2 f(m)/\partial q^2
    \end{bmatrix}\nonumber \\
    &=
    \begin{bmatrix}
        1-(A+B) & K (A-B) \\
    K (A-B) & K-K^2 (A+B)
    \end{bmatrix}
\end{align}

with $A=\frac{1}{\sqrt{2 \pi R^2}}\exp (-(m-z)^2/2 R^2)$ and $B=\frac{1}{\sqrt{2 \pi R^2}}\exp (-(m+z)^2/2 R^2)$. 

A fixed point is a local minima if:  a)  the determinant  ($D$) of the above matrix,  is $D \geq 0$  and b) the first element, $1-A-B \geq 0$. The determinant $D$ is given by 
\begin{eqnarray}
D &=& K^2  \left(\frac{1}{K}+c^2\right)-c K^2 \left(\frac{1}{K}+1\right) \cosh(m z/R^2)
\end{eqnarray}
where $c= \sqrt{\frac{2}{\pi R^2}} e^{(-m^2-z^2)/2 R^2}$.

The transition where the steady state changes from one fixed point in  $(m,q)$ to another under Glauber Dynamics for a given  $K$ and $\Delta$ occurs when  $1-A-B=0$  and $D=0$. 

The $1-A-B=0$ gives the condition:
\begin{equation}
    c  \cosh (m z/R^2)=1
\end{equation}
Similary, the $D=0$ results in the condition:
\begin{equation}
c \cosh (mz/R^2) =\frac{1+K c^2}{1+K}
\end{equation}
The two  conditions together are satisfied when $c_{c}=1$. For example, for the fixed point with $m=0$ the condition $c_{c}=1$  is equivalent to 
\begin{equation}
\sqrt{\frac{2}{\pi R_{c}^2}} e^{-z_{c}^2/2 R_{c}^2}=1
\label{fpz}
\end{equation}
This equation is same as the Eq. \ref{bcmzerofp}, the equation for transition under Glauber dynamics for $m=0$ initial state with $\Delta$ being replaced by $z$ with a crucial difference: In the case of RFBEGM, $z$ is not fixed for a given $K$ and $\Delta$ and changes during the evolution depending on the value of $q$. 

For $K>0$, when $c \leq 1$,  $m=0$  is the  local  minima  that becomes a local maxima for $c >1$. Hence the solution of the above equation gives the phase boundary in $z-R$ plane under Glauber dynamics in this case. If in the steady state $z>0$, then it is like RFIM and if $z<0$ then it behaves similarly to RFBCM with $\Delta>0$.  Since Eq. \ref{fpz} is same as Eq. \ref{bcmzerofp} with $\Delta$ replaced by $z$, the phase boundary is the same as $m=0$ phase boundary in Fig. \ref{fig2}.

For $K<0$, the RFBEGM behaves differently than the RFBCM. For  $z < 0$, while $D \geq 0$ for $c \geq 1$, $1-A-B \geq 0$ when  $c \leq 1$. Hence the system cannot satisfy both the conditions simultaneously except for $R=0$. Also since 
$\partial^2 f/\partial q^2 < 0$ for $K<0$, there is no transition in  this  case for $z<0$. 

For $z=0$  the free energy functional $f(m,q)$ has  a simpler  expression. We hence study $z=0$ below separately.
%While for $z<0$ to decide the the nature of transition and hence the location of the tricritical point, one will have to undertake higher order stability analysis (till foruth order derivatives of $f(m,q)$), for $z=0$

\subsubsection{$z=K q- \Delta =0$}
The free energy functional for $z=0$ takes a simpler form and is given by
\begin{align}
   f(m,q) &= \frac{m^2}{2}+\frac{K q^2}{2}-m~\text{erf}\left(\frac{m}{\sqrt{2} R}\right)- \frac{R}{\sqrt{2 \pi}} \text{exp}\left(-\frac{m^2}{2 R^2}\right)
\end{align}
The value of $q$ is constrained by the $z=0$ condition. It can take only two values: $q=1$ if $\Delta=K$ and $q=\Delta/K$ if $\Delta > K$. These two values do not correspond to the fixed points of $f(m,q)$ which is trivially $q=0$. The fixed point equation for $m$ does not depend on $q$ and is given by
\begin{equation}
    m =\text{erf} \left(\frac{m}{\sqrt{2} R}\right)
\end{equation}
Since $f(m,q)$ depends on $q$, and $q$ does not change the fixed points of $f(m,q)$, it is worth considering both the values of $q$.

For $q=1$, $f(m,q)$ is
\begin{align}
   f(m,1) &= \frac{m^2}{2}+\frac{K}{2}-m~\text{erf}\left(\frac{m}{\sqrt{2} R}\right)- \frac{R}{\sqrt{2 \pi}} \text{exp}\left(-\frac{m^2}{2 R^2}\right)
\end{align}
and for $q=\Delta/K$ it is
\begin{align}
   f\left(m,\frac{\Delta}{K}\right) &= \frac{m^2}{2}+\frac{\Delta^2}{2  K}-m~\text{erf}\left(\frac{m}{\sqrt{2} R}\right)- \frac{R}{\sqrt{2 \pi}} \text{exp}\left(-\frac{m^2}{2 R^2}\right)
\end{align}
The $q=\Delta/K$ minimises $f(m,q)$ when $\Delta^2 < K^2$, with $\Delta> K$. This is possible only when both $\Delta$ and $K$ are negative. The critical value of $R$ is $R_c =\sqrt{\frac{2}{\pi}}$, which matches with the $R_c$ obtained by taking $c=1$ at $(z,m)=(0,0)$ in the above analysis. The other possibility $q=1$, occurs only for $\Delta=K$. The $R_c$ for this state is also the same, i.e $R_c=\sqrt{\frac{2}{\pi}}$.

\begin{figure*}[htbp]
\centering
\includegraphics[width=0.6\hsize]{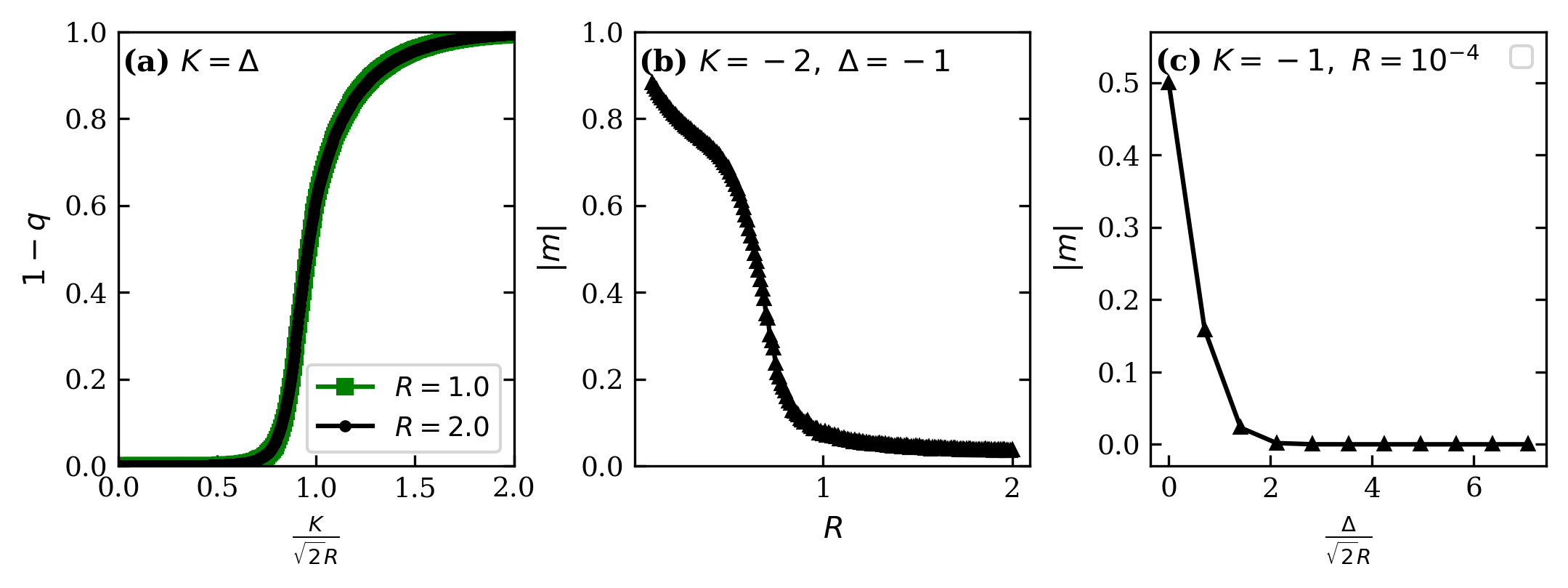}
\caption{Plots for RFBEGM Glauber steady state  with $K  \leq \Delta$ for: a) For  $K = \Delta$ we plot $1-q$ as a function of $K/(\sqrt{ 2} R)$ and find a data collapse for $R=1$ and $2$ verifying Eq. \ref{Kequalsd}. The transition is at $K/(\sqrt{ 2} R)=\sqrt{\pi/4}\approx 0.88$; b) For $K < \Delta < 0$, we take $K = -2, \Delta = -1$ and plot $|m|$ as a function of $R$. It starts to decay first suggesting no transition, but once $q<\Delta/K$, the model shows a crossover and has a transition at $R_c=\sqrt{2/\pi}$; c) For $K < \Delta$ with $K < 0$ and $\Delta > 0$, we take $K = -1$ and verify that there is no transition even at any value of $R$ by plotting  $|m|$ as a function of $\Delta (>0)$ for a value of $R = 10^{-4}$. All the plots are obtained for a system of size $N = 1000$, averaged over $1000$ realizations of random field distributions.}
\label{fig4}
\end{figure*}

We hence can now discuss the phase transitions for Glauber dynamics. For $K>\Delta$ or $z >0$ the behavior is similar to RFIM and hence we will not discuss that here. For $K  \leq \Delta$, the behavior of RFBEGM can be separated based on the values of $K$ and $\Delta$ and the following four main situations arise:

\begin{enumerate}
\item $K=\Delta$. In this case $z= K(q-1)$. 

For negative $K$, $q=1$ along $K=\Delta$ line and the model has the same phase behaviour as the standard RFIM.

On the other hand, for positive $K$, $z<0$ is possible only when $q <1$. Taking $m=0$ in the fixed point equation of $q$ we get
\begin{equation}
    1-q =\text{erf} \left( \frac{K (1-q)}{\sqrt{2}  R}\right)
    \label{Kequalsd}
\end{equation}
Hence for $K \leq K_c= \sqrt{\frac{\pi R^2}{2}}$, $1-q=0$, while for $K>K_c$ , $(1-q)$ can also be non zero.  Hence, for $K\leq K_c$, the system is like RFIM with $q=1$ and for $K>K_c$ it has $z<0$ and  $q \leq 1$. The transition in $m$ occurs at $R_c=\sqrt{\frac{2}{\pi}}$. For $R<R_c$ the transition is first-order and for $R>R_c$ it is second-order. In Fig. \ref{fig4}(a) we plot $1-q$ vs  $K/(\sqrt{2} R)$ from simulations for $N=1000$ spins, which agree very well  with the theoretical predictions.

\item $K < \Delta$ with $K<0$ and $\Delta <0$. 

In this case, $z$ can take any value. If $q> \Delta/K$, then $z>0$; if $q=\Delta/K$ then $z=0$ and if $q< \Delta/K$ then $z<0$. But at  $R=0$ the system is always in $|m|=1$ state. This corresponds to $z<0$. As a result the Glauber steady state show continuous decrease in the values of $m$ and $q$  for very small $R$. This is expected as when  $z<0$ there is no transition. 
But once $q \leq \Delta/K$, the  $z$ becomes greater than $0$. If this occurs at $R<\sqrt{\frac{2}{\pi}}$, then the system has a continuous order-disorder phase transition at $R=R_c=\sqrt{\frac{2}{\pi}}$. This is illustrated  via simulations in Fig. \ref{fig4}(b). Hence we find that the model undergoes a crossover from a model with no transition to a model with $R_c=\sqrt{\frac{2}{\pi}}$ in this region.

%A system prepared in  $m=0$ state with $q \geq \Delta/K$ undergoes transition at $R_c=\sqrt{\frac{2}{\pi}}$ as this is the $R_c$ for $z \geq 0$. On the other hand a system prepared in an initial state with $m=0$ and $q <\Delta/K$ has $R_c=0$.

\item $K <\Delta$ with $K<0$ and $\Delta >0$: This  corresponds to $z<0$ and hence  again there is no transition. This is verified  in the simulations  as shown in Fig. \ref{fig4} (c) for $K=-1$.

\item $K <\Delta$ with $K > 0$ and $\Delta >0$:  This again corresponds to $z<0$ lying  between $-\Delta$ and $K-\Delta$. This though  differs from the $K<0$,  $z<0$ cases discussed above as in this case  due to positive $\Delta$ the $0$ spin is preferred giving rise to $m=0$ as a preferred state. The RFBEGM behaves similar to RFBCM with $\Delta>0$ in this regime. We will see in Sec.  \ref{sec7}, as a result of $0$ spins being favored, the model exhibits double hysteresis in the presence of external magnetic field where two loops are separated by a state with predominantly  $0$ spins.
\end{enumerate}
For the equilibrium phase diagram, we need to find the condition of tricriticality for the two variable $f(m,q)$ by looking at higher order derivatives for a given value of $K$ and $\Delta$ in general. Although for some values of $K,\Delta$  we can predict the phase diagram by using the analogy between $z$ in RFBEGM and $-\Delta$ in RFBCM. For example, in terms of $z$, the condition of tricriticality would become $z_{tcp}= -0.483$. This then gives $\Delta_{tcp}= Kq-z_{tcp}$. 
For $K<0$ and $\Delta>0$ with $\Delta>K$, this condition is not satisfied unless $q$ is very small and hence in that quadrant equilibrium and non-equilibrium steady  states behave similarly: there is no transition at any value of $R$.

\section{Hysteresis and phase transitions in the presence of the external field H}
\label{sec7}
The random field spin models with a first-order equilibrium transition usually exhibit hysteresis under Glauber dynamics as the final state  reached depends on the initial state. For example, RFIM for $R<R_c=\sqrt{2/\pi}$ exhibits hysteresis in the $m-H$ plot that  look like Fig.  \ref{fig5} (a). We do not discuss the hysteresis in RFIM explicitly as it has been extensively studied \cite{sethna} and is similar to $\Delta <0$ hysteresis of RFBCM.

The shape of hysteresis is used to understand the properties of the material \cite{dhar}. Also the coercive field, the $H$ at which the magnetization becomes non zero is an important way to characterize  materials \cite{coercive}. We find that RFBEGM shows a variety of hysteresis behavior as a function of $K$ and $\Delta$. The coercive field also depends on $K$ and $\Delta$. Though the hysteresis area changes with $R$, the shape of the hysteresis loop  is determined by the behavior of the model at $R=0$ under Glauber dynamics. We hence study the hysteresis for $R=0$ and the effect of $R$ on the $R=0$ behavior in this section. We study hysteresis between the two saturated states $m=\pm 1$.
\subsection{RFBCM in external magnetic field}
\subsubsection{$R=0$}
Let us first consider the Blume-Capel model without disorder, in the presence of external field. The energy per spin is given by:
\begin{equation}
E =-\frac{m^2}{2} +\Delta q -H m  
\end{equation}
For $H=0$, in the case of equilibrium there is a first-order transition at $\Delta=0.5$, where $(m,q)$ changes from $(\pm1,1)$ to $(0,0)$. For $H \neq 0$, the equilibrium first-order transition shifts to $\Delta_{\pm}=0.5 \pm H$. 

\begin{figure*}[htbp]
\centering
\includegraphics[width=0.4\hsize]{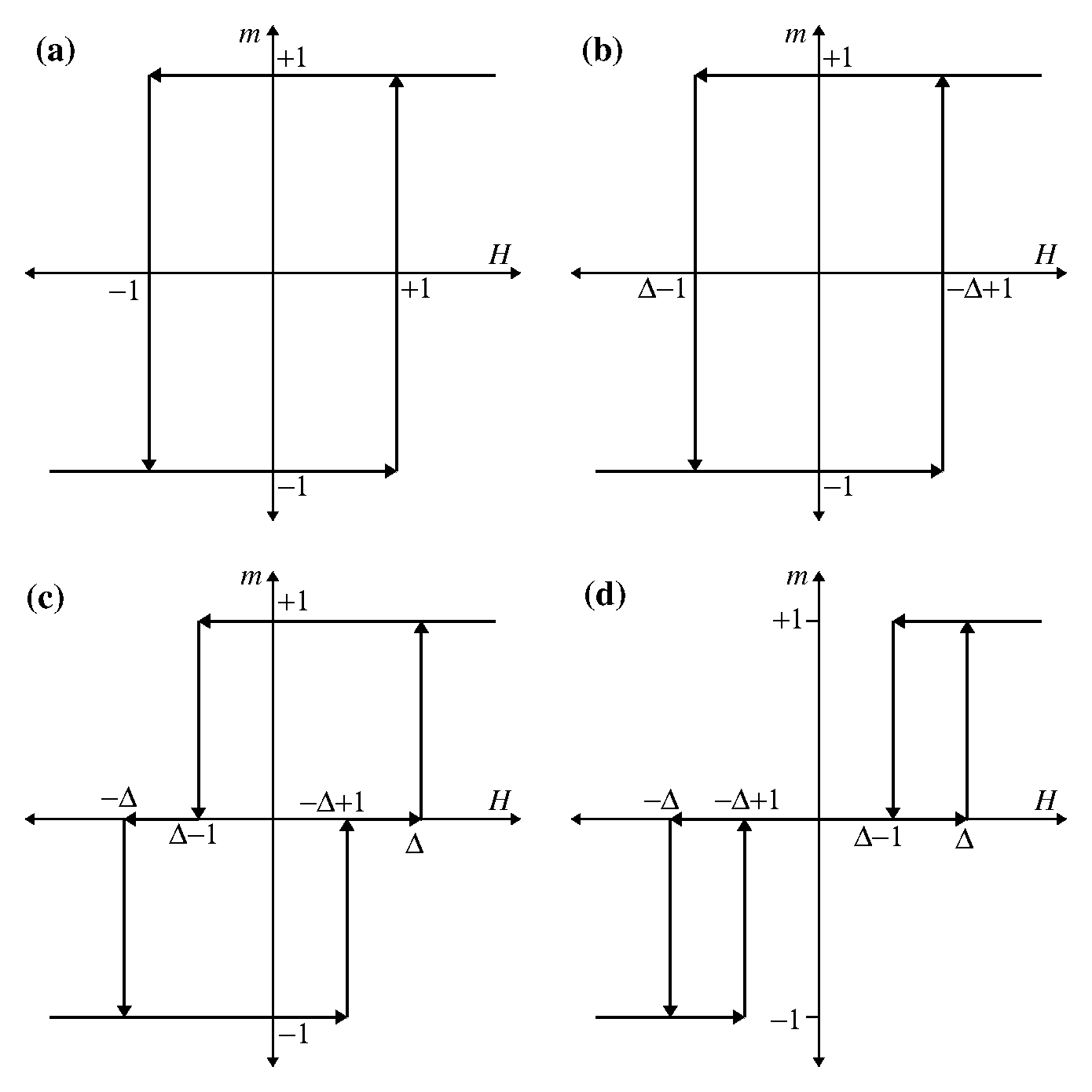}
\caption{Different shapes of hysteresis loops in $m-H$ plane for Blume-Capel model that depend on the value of the crystal field $\Delta$: a) $\Delta \leq 0$: rectangular with the coercive field ($H_c$) independent of the value of $\Delta$; b) $0 < \Delta \leq 0.5$: Rectangular with ; $H_c = \pm (\Delta-1)$ c) $0.5 < \Delta < 1 $: Wasp-waisted with $H_c =\pm (\Delta-1)$ and $\pm \Delta$ and d) $\Delta >1$: Double rectangular with $H_c = \pm (\Delta-1)$ and $\pm \Delta$.}
\label{fig5}  
\end{figure*}

On the other hand, for evolution under Glauber dynamics, for $R=0$, the steady state reached depends on the initial state of the system. This hence introduces hysteresis in the presence of the external magnetic field  $H$. Depending on the value of $\Delta$, the form  of  the hysteresis  curve changes.  We study  hysteresis  starting with a  state in which  $s_i=-1$ $\forall i$. The external field H is increased quasi-statically to generate a hysteresis loop. The evolution can be understood from Eqs. \ref{lf1} and \ref{lf2}. For $\Delta \leq 0$, $L_2$ plays no role, and $m$ jumps from $-1$ to $+1$ as a function of $H$ when $L_1=0$. This results in a rectangular hysteresis loop with jumps at the coercive fields $H_c=\pm  1$ just like the Ising model as shown in Fig. \ref{fig5}(a). 

For $\Delta >0$ , when $L_2=L_1$, the spins first flip to $0$. For $\Delta <1/2$, the $0$ spin state is not stable and without any further increase in $H$, all the spins flip to $+1$. Hence for $\Delta <1/2$ the shape of the hysteresis loop is still rectangular but the value of the coercive field becomes a function of $\Delta$  and is equal to $H_c=\pm (1-\Delta)$. For $\Delta>1/2$, the $0$ spin state becomes stable between $1-\Delta$ and $\Delta$ during the quasi-static evolution starting with $m=-1$ state and between $\Delta-1$ and $-\Delta$ during the quasi-static evolution starting with $m=+1$ state. Hence for $\Delta > 1/2$, there are four first-order transitions in the hysteresis. For $\Delta >1$, $m-H$ curve separates into two well separated hysteresis loops giving rise to a double hysteresis loop with four first-order transitions as shown in Fig. \ref{fig5}(d).

\subsubsection{with disorder}
For $R \neq 0$, it is not possible to follow the deterministic evolution to find the steady state as done in the previous section. But the values of the coercive field where the phase transition occurs as a function of $H$ can be obtained by extending the treatment of Sec. \ref{sec5}(b). For $H \neq 0$
we get the condition for a phase transition for an initial state prepared in the basin of a fixed point $m$ by
\begin{align}
1&=\sqrt{\frac{1}{2 \pi R^2}} \left(e^{-(m+H-\Delta)^2/2 R^2}+e^{-(m+H+\Delta)^2/2 R^2}\right)\\
1&=\sqrt{\frac{1}{2 \pi R^2}} e^{-(m+H-\Delta)^2/2 R^2} \left(1+e^{-2  \Delta(m+H)/R^2}\right)
\label{bcfpH}
\end{align}
For example, for $R=0$ the above condition for the phase transition for $m+H >0$ comes out to be
\begin{equation}
m+H-\Delta=0
\end{equation}
Hence starting from $m=+1$ state, there is a transition to $m=0$ state at $H_c= -1+\Delta$ in the backward loop.
Similarly for $m+H<0$ the transition occurs at 
\begin{equation}
m+H+\Delta=0
\end{equation}
Which implies that starting from $m=-1$ state, there is a transition to $m=0$ state at $H_c= 1-\Delta$ in the forward loop.

Depending on the value of $\Delta$, another transition occurs at $\Delta$ and $-\Delta$ respectively for backward and forward loops from $m=0$ state. These results match with the analysis done in the previous section for $R=0$ directly from the dynamical equations.

\begin{figure*}[htbp]
\centering
\includegraphics[width=0.6\hsize]{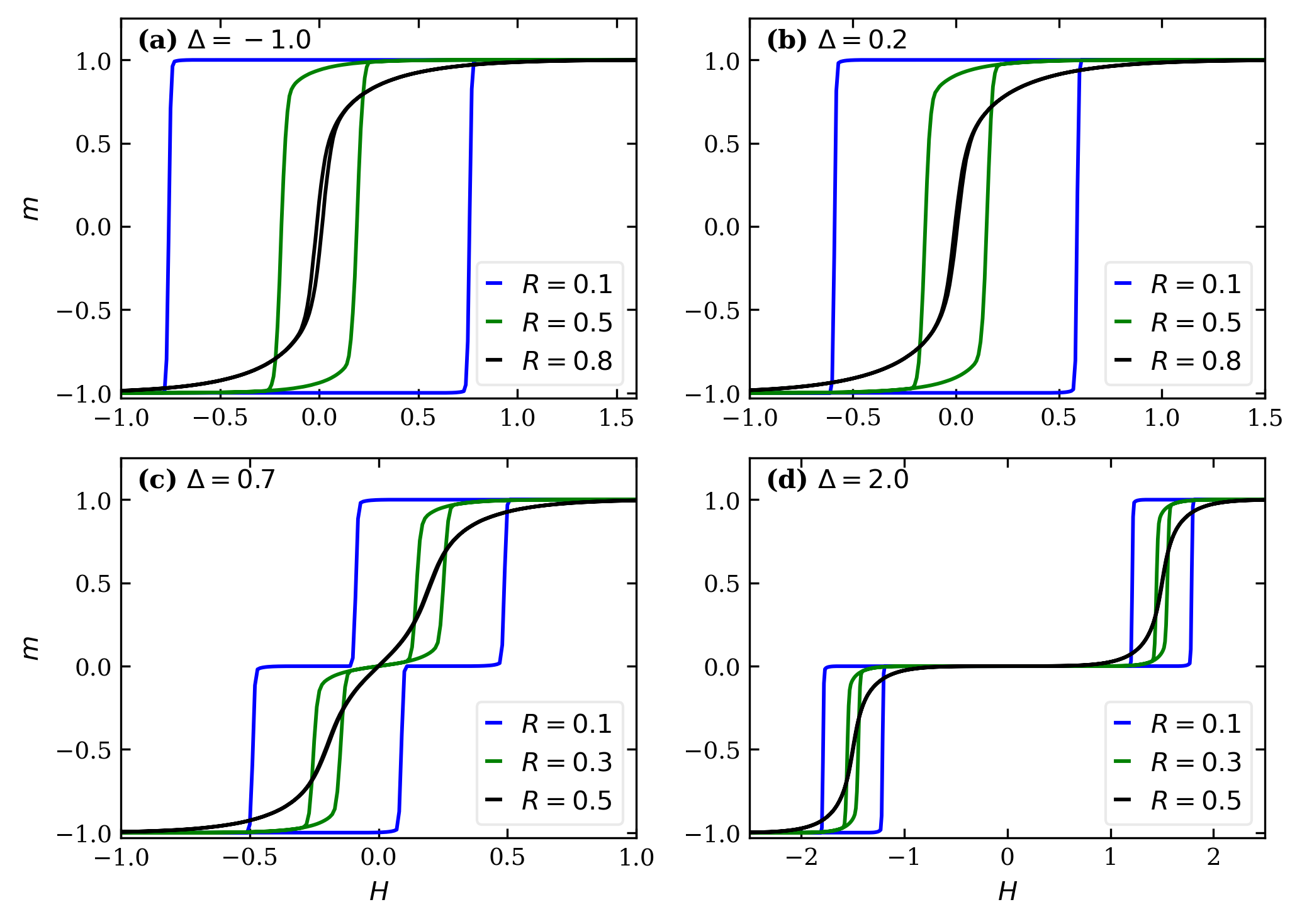}
\caption{The $m-H$ plots in the presence of disorder are plotted for few representative values of $\Delta$ corresponding to four regions in Fig. \ref{fig5}. The shape of the plot for $R=0$ is retained till the hysteresis vanishes at $R = R_c$. The data for all the plots are obtained for a system of size $N = 1000$, after averaging over $1000$ realizations of random field distributions for each value of $H$.}
\label{fig6}
\end{figure*}

For $R \neq 0$, the hysteresis loops stay first-order for $R<R_{c}$ with the value of coercive field at the transition given by the solution of Eq. \ref{bcfpH} for a given $\Delta,R$ and the initial state. This can be obtained by solving the equation numerically. As $R$ increases the area under the hysteresis loops shrink and there exist a value of $R=R_c$  beyond which there are no solutions to the Eq. \ref{bcfpH} and hence there is no hysteresis for $R> R_{c}$. The shape of the hysteresis loops for $R<R_c$ is the similar to the $R=0$ hysteresis loops. This is illustrated in Fig. \ref{fig6} where hysteresis loops as a function of $R$ are plotted for representative $\Delta$ values.

\subsection{RFBEGM in external magnetic field}

\subsubsection{R=0}
\begin{figure*}[htbp]
\centering
\includegraphics[width=0.6\hsize]{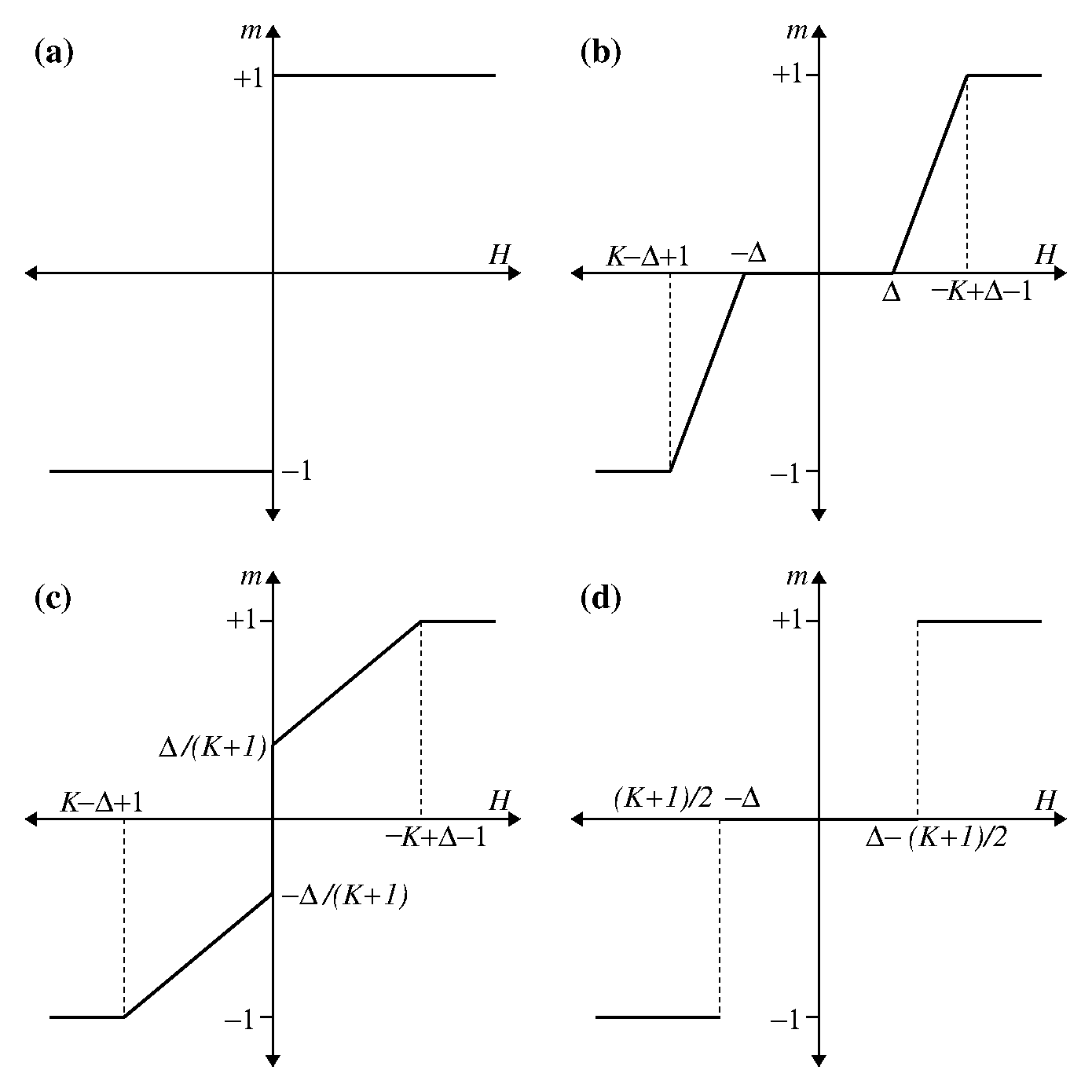}
\caption{Different shapes of $m-H$ plots for the equilibrium  Blume-Emery-Griffith model that depend on the value of $K$ and $\Delta$: a) $K+1< 0$, $\Delta < K+1$ or $K + 1 > 0, 2\Delta - K - 1 < 0$: b) $K+1<0$, $\Delta > 0$: c) $K+1<0$, $K+1<\Delta < 0$: The value of $0<|m|< 1$ in the interval $H\in(K-\Delta+1, -K+\Delta-1)$ and d) $K+1>0,\ 2\Delta-K-1 >0$}
\label{fig7}
\end{figure*}

\begin{figure*}[htbp]
\centering
\includegraphics[width=0.7\hsize]{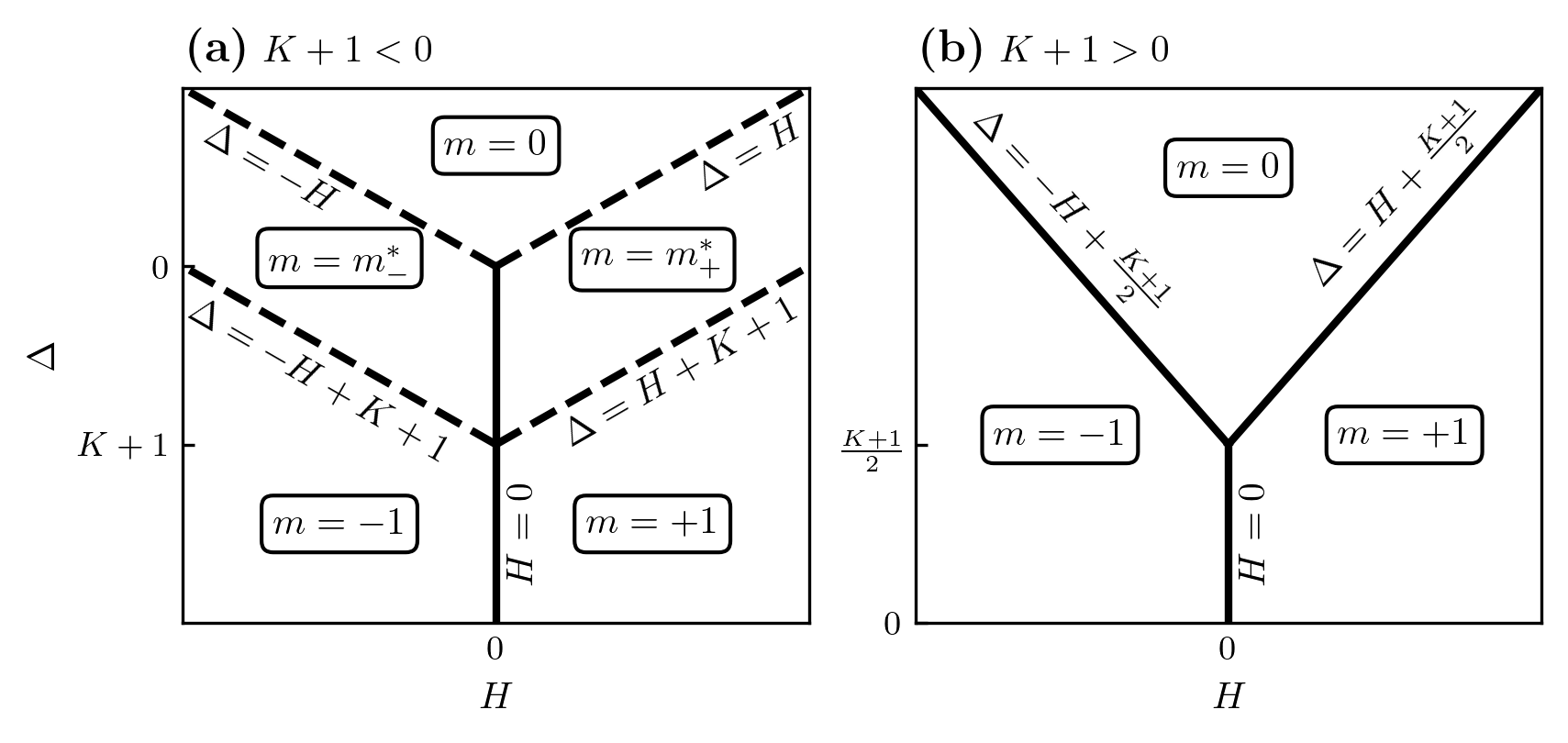}
\caption{The equilibrium phase diagrams of the Blume-Emery-Griffiths model are given in this figure. The regions of different phases are labeled by their respective values of equilibrium magnetisation, $m$ (given in boxes). Here, $m^*_{\pm}=\frac{\pm\Delta-H}{K+1}$. The solid and dashed lines denote discontinuous and continuous phase transitions respectively.}
\label{fig8}
\end{figure*}

The $m-H$ plots for RFBEGM change non-trivially with values of $K$  and  $\Delta$. The  energy is given by
\begin{equation}
E =-\frac{m^2}{2}-\frac{K}{2} q^2  +\Delta q -H m  
\end{equation}
Since $|m| \leq  q$, for a given $q$, $E$ is minimum when $|m|= q$. Hence we can write 
\begin{equation}
E =-\frac{K+1}{2} q^2  +(\Delta-H) q 
\end{equation}
Depending on  $(K,\Delta,H)$, the above equation has three fixed points $q=0,1$ and $q^*$, where $q^*=\frac{\Delta-H}{K+1}$. For  $K+1>0$, only $q=0$ and  $1$  minimise the energy $E$ and for $K+1<0$, all three fixed point values  of $q$ can exist. Different  possible equilibrium $m-H$ plots are shown in Fig. \ref{fig7}. Depending on the values of $K$ and $\Delta$, RFBEGM also has a continuous transitions even at $T=0$ as shown in Fig. \ref{fig7} (b) and (c). As a result there are two different $\Delta-H$ phase diagrams corresponding to $K<-1$ and $K>-1$ as shown in Fig. \ref{fig8}.
%This can be summarised in the phase  diagrams in the $\Delta-H$ plane as shown in Fig. \ref{fig8}

% \begin{figure*}[htp]
% 	\centering
% 	\begin{subfigure}[b]{0.4\textwidth} % Adjust the width as needed
% 		\centering
% 		\includegraphics[width=\textwidth]{PD_equi_BEGM_1.png} % Replace with your image file
% 		\subcaption{$K + 1 < 0$} % Add a meaningful caption
% 	\end{subfigure}
% 	\hfill
% 	\begin{subfigure}[b]{0.36\textwidth} % Adjust the width as needed
% 		\centering
% 		\includegraphics[width=\textwidth]{PD_equi_BEGM_2.png} % Replace with your image file
% 		\subcaption{$K + 1 > 0$} % Add a meaningful caption
% 	\end{subfigure}
% 	\caption{The equilibrium phase diagrams of the BEGM are given in this figure. The regions of different phases are labeled by their respective values of equilibrium magnetisation, $m$ (given in boxes). Here, $m^*_-,\ m^*_+\neq 0, \pm 1$. The solid and dashed lines denote discontinuous and continuous phase transitions.}
% 	\label{fig8} % Optional: Add a label for referencing
% \end{figure*}

For Glauber dynamics, the  model exhibits hysteresis. The shape of the  hysteresis plots non-trivially change with values of $K$ and $\Delta$ even in the absence of random fields. These can be derived from Eq. $2$ and $3$. We derive the $m-H$ curves for $s_i=-1$ $\forall i$ initial condition for different cases  below.
\begin{figure*}[htbp]
\centering
\includegraphics[width=0.8\hsize]{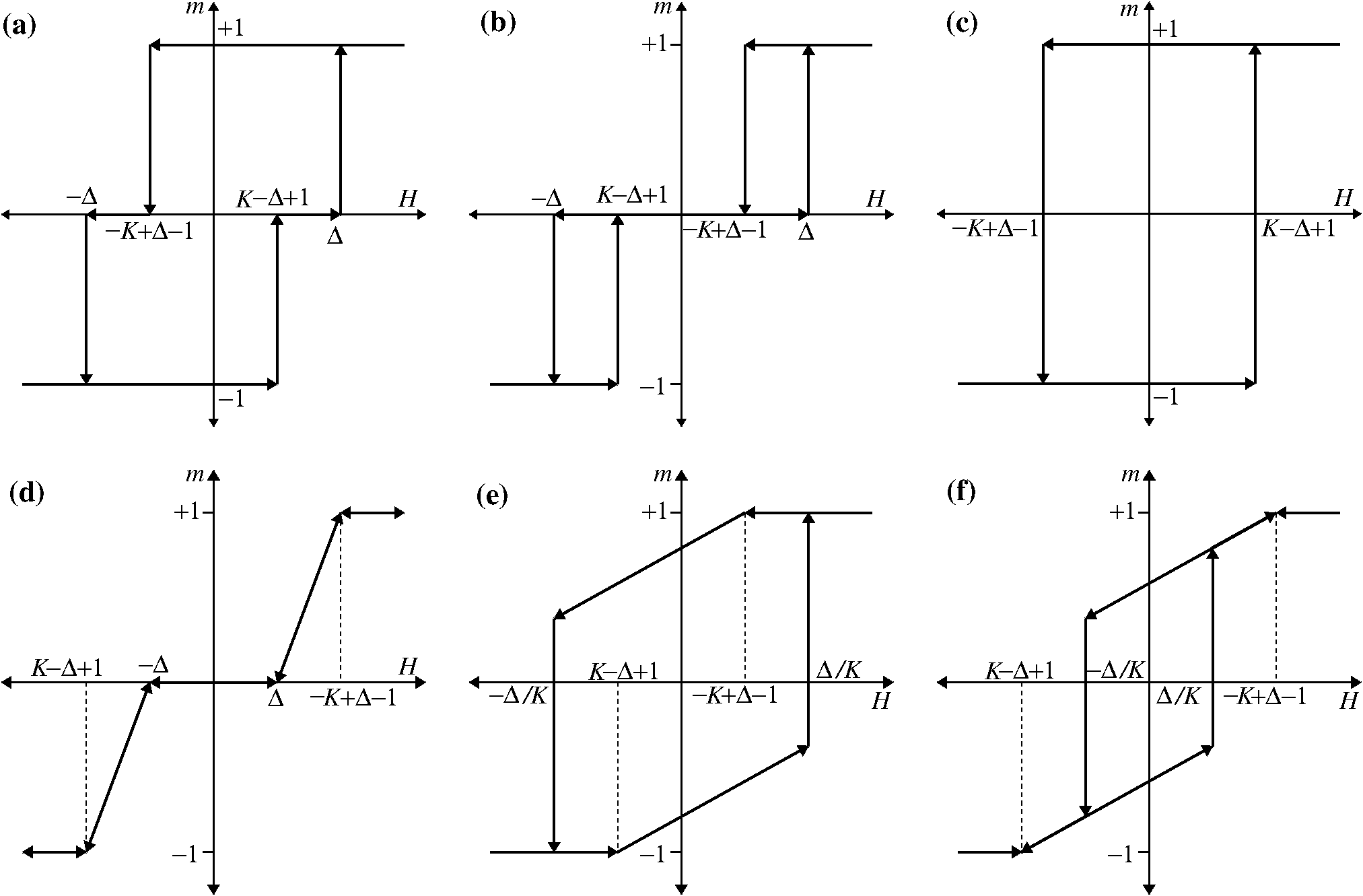}
\caption{ Different shapes of hysteresis loops in $m-H$ plane for Blume-Emery-Griffith model for Glauber dynamics: a) $|K|<1,\ (K+1)/2 < \Delta < K+1$   (or)     $K>1,\ K< \Delta<K+1$: Wasp-waisted;  b) $K > -1,\  \Delta > K+1$: Double rectangular; c) $K < \Delta,\ |K| < 1,\ \Delta > 0,\ K > 2\Delta-1$ or $K < \Delta,\ K \in (-1, 0), \Delta < 0$: Rectangular ; d) $K < \Delta,\ K < -1,\ \Delta > 0$: There is no hysteresis. $m$ is a continuous function of $H$; e) $K < -1,\ \Delta \in (K,\frac{K(K+1)}{K-1})$: Hexagonal; f)$K < -1, \frac{K(K+1)}{K-1}<\Delta <0$: Parallelogram.}
\label{fig9}
\end{figure*}

\begin{figure*}[htbp]
\centering
\includegraphics[width=0.9\hsize]{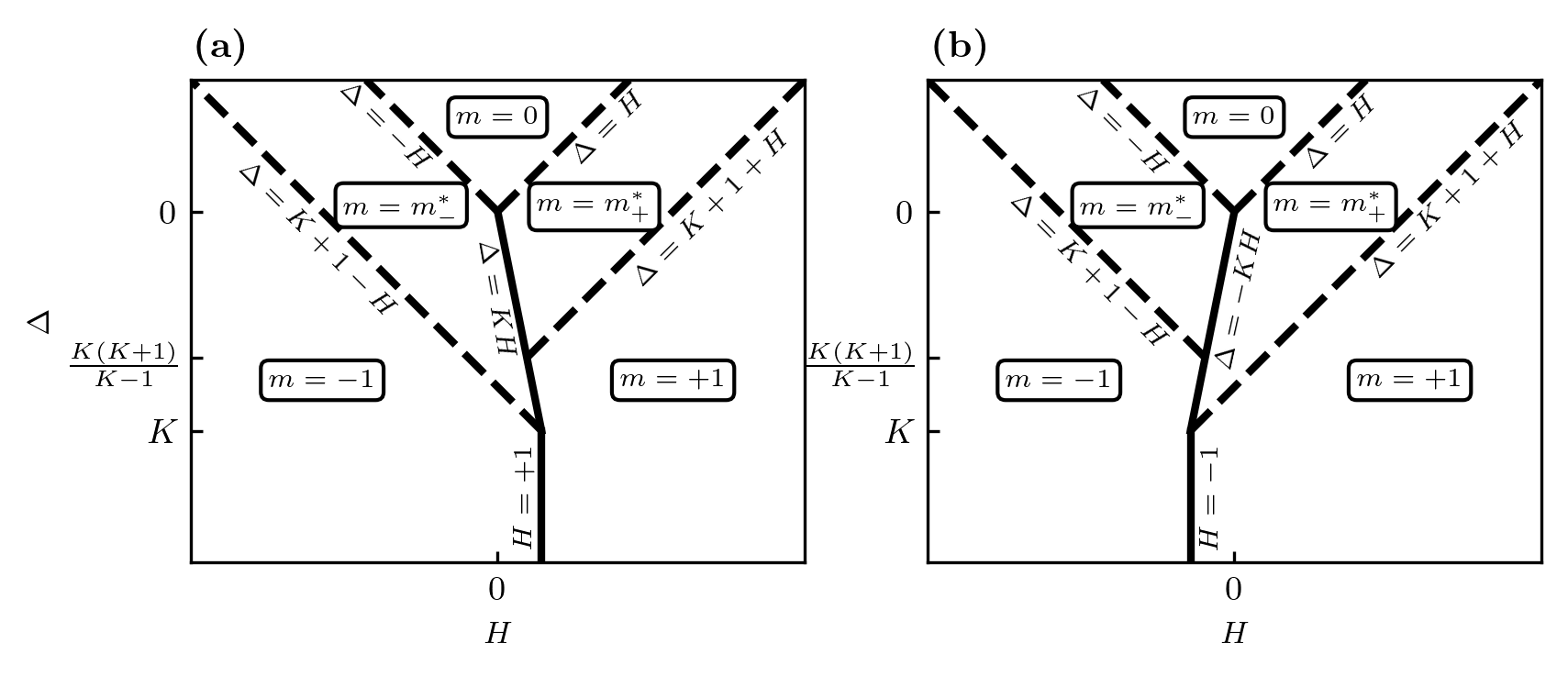}
    \caption{Phase diagram for a) $m=-1$ and b) $m=+1$ saturated states as initial states for Glauber dynamics steady state of Blume-Emery-Griffiths model for $K<-1$. Solid and dashed lines represent first and continuous phase transitions respectively. Here, $m^*_{\pm}=\frac{\pm\Delta-H}{K+1}$.}
\label{fig10}
\end{figure*}

\begin{enumerate}
\item $K> \Delta$: In this case  $L_2(s_k) > 0$ through out the dynamics and hence the steady state is same as that of the Ising model. The  coercive field for forward  and backward loops  is $H_c=1$ and $H_c=-1$ respectively as  shown in Fig.  \ref{fig5} (a).

\item $K< \Delta$  with $|K|<1$ , $\Delta > 0$ and $K<2\Delta-1$: Initially, since $s_i=-1$ $\forall i$,

$$L_1(s_k) = -\frac{1}{N} (N-1) + H ;$$ $$L_2(s_k) = \frac{1 + K}{2N} + \frac{K}{N} ( N-1 )  - \Delta $$
The  spin-flip of $-1$ to $0$ happens when $L_1(s_k) =L_2(s_k)$ at a coercive field of $H = H_c^{(1)}$, provided $L_2(s_k)< 0$. As a spin-flips from $-1$ to $0$ the $L_1(s_k)$ increases by $1/N$ and $L_2(s_k)$ increases or decreases by $|K|/N$ as long as $L_2(s_k)<0$. Since $|K| < 1$, spin-flip of $-1$ to $0$ is preferred, as long as $L_2(s_k)< 0$. But, since $\Delta > 0$, $L_2(s_k)< 0$ is always satisfied for all values of $q$. Also, since, $K<2\Delta-1$, $L_1(s_k) < -L_2(s_k)$. Thus, at $H = H_c^{(1)}$ all spins flip to $0$. At this point  in dynamics, $s_i = 0\ \  \forall i$ and hence
$$L_1(s_k) = H;~L_2(s_k) = \frac{1 + K}{2N} - \Delta $$
The  spins will flip  to $+1$ when $L_1(s_k) = -L_2(s_k)$, where all the spins flip to $+1$. Thus, in the limit $N \rightarrow \infty$, we have a second coercive field  $H_c^{(2)}$  where all the spins flip to $+1$.
In the limit $N \rightarrow \infty$, we have
\begin{equation}
H_c^{(1)} = K - \Delta +1;\ H_c^{(2)} = \Delta
\end{equation}

We now decrease the  magnetic field quasi-statically after  all the spins  become $+1$  to obtain the backward $m-H$ curve. At this moment

$$L_1(s_k) = \frac{1}{N} (N-1) + H;$$  $$~L_2(s_k) = \frac{1 + K}{2N} + \frac{K}{N} \bigg( N-1 \bigg)  - \Delta $$
Similar  arguments as above then gives:
\begin{equation}
H_c^{(1)} = -K + \Delta-1;~H_c^{(2)} = -\Delta
\end{equation}
Depending on the value of $K$ and $\Delta$  the hysteresis plot takes shapes as shown in Fig. \ref{fig9} (a) and (b).

\item $K < \Delta$ with $|K| < 1$, $\Delta > 0$ and $K>2\Delta-1$: This is similar to the  previous scenario, except for the  fact that since now $K> 2 \Delta-1$, after $x = 2\frac{\Delta-K}{1-K}$ fraction of spins flip to $0$,  $L_1(s_k) = -L_2(s_k)$. As a result all the spins flip to $+1$.  The hysteresis plots looks like that for the Ising model (Fig.\ref{fig9} (c)) with forward  and backward coercive fields given by
$$H_c^{(1)} = K - \Delta +1; H_c^{(2)} = -K + \Delta -1$$
respectively.
\item $K < \Delta$ with $|K| < 1$ and $\Delta < 0$: Similar arguments as above follow and we have the hysteresis plots have similar shape(Fig. \ref{fig9} (c)  with the same values of the coercive fields:
$$H_c^{(1)} = K - \Delta +1;H_c^{(2)} = -K + \Delta -1$$

\item $K < \Delta$ with $K > 1$

For the initial state: $s_i = -1 \forall i$ we have
$$L_1(s_k) = -\frac{1}{N} (N-1) + H;$$  $$~L_2(s_k) = \frac{1 + K}{2N} + \frac{K}{N} \bigg( N-1 \bigg)  - \Delta $$
Transition from the state of all spins $s_i = -1$ to the state of all spins $s_i = 0$ happens at $H = H_c^{(1)}$ when $L_1(s_k) = L_2(s_k)$. Transition from the state of all spins $s_i = 0$ to the state of all spins $s_i = +1$ happens at $H = H_c^{(2)}$ when $L_1(s_k) = -L_2(s_k)$. The values of the forward and backward coercive fields are

$$H_c^{(1)} = K-\Delta+1 ; H_c^{(2)} = \Delta$$

Similarly when we start with the Initial state: $s_i = +1 \forall i$, transition from the state of all spins $s_i = 1$ to the state of all spins $s_i = 0$ happens at $H = H_c^{(1)}$ when $L_1(s_k) = L_2(s_k)$. Transition from the state of all spins $s_i = 0$ to the state of all spins $s_i = +1$ happens at $H = H_c^{(2)}$ when $L_1(s_k) = -L_2(s_k)$,  giving
 $$H_c^{(1)} = -K+\Delta-1;H_c^{(2)} = -\Delta$$
This results in hysteresis loops as shown in Figs. \ref{fig9} (a) and (b).

\item $K < \Delta$ with $K < -1$ and $\Delta > 0$: This is a non-trivial case where there is no hysteresis. Frustration due to negative $K$ wipes out the first-order transition. This can be seen as follows:

Starting with all  spins  $-1$, the  first spin-flip to a $0$ state happens when $L_1(s_k) = L_2(s_k)$ at
$H =K - \Delta +1$. The system reaches a steady state at the end of a single flip. All the 
spins get flipped one by one  to $0$ with increments of $\delta H = \frac{|K|-1} {N}$  in $H$ untill all  spins become $0$  and $m=0$  , this occurs at  $H_c^{(1)} = H+N((|K|-1)/N)$. Thus $$H_c^{(1)} = -\Delta$$. 

The values of $L_1$ and  $L_2$ are now
$$L_1(s_k) = H;~L_2(s_k) = \frac{1 + K}{2N}  - \Delta $$

The next flip happens $L_1(s_k) = -L_2(s_k)$. This  happens at   
$$H_c^{(2)} = \Delta$$

The system again approaches $m=+1$ state with increasing magnetisation with single spin-flips with increments of $\delta H = \frac{|K|-1} {N}$. This, occurs at $H  = \Delta - K - 1$.

The resulting $m-H$ plot is shown Fig. \ref{fig9} (d). In this case all the transitions are continuous as oppose to first-order transitions for cases discussed above and in Blume-Capel model due to the frustration induced by $K$.

\item  $K < \Delta$ with $K < -1$ and $\Delta < 0$
Starting with $s_i=-1$ $\forall i$, the first spin-flip to a $0$ state happens when $L_1(s_k) =L_2(s_k)$ at  $H= K-\Delta+1  = H_c $. After one spin-flip the system reaches a steady state and the system requires $\delta H = \frac{|K|-1} {N}$ for the next spin to flip, after which another steady state is reached. As long as $L_2(s_k) <0$, this process continues. After $x N$ spins flip to $0$, $L_2(s_k) = 0$. The value of $x$ when $L_2(s_k)=0$ is $x = \frac{K-\Delta}{K}$.  Thus 
$$H_c^{(1)} = H_c + xN\delta H= \Delta/K$$
The steady state, thus reached, has $xN$ zero spins and $(1-x)N$ spins of $-1$. Upon an increase in $H$ by $\delta H$, all the $-1$ spins become unstable and flip to $+1$. In the limit $N\rightarrow\infty$, the local fields of the zero spins are given by: $L_1(s_k)=\frac{2\Delta}{K}>0$ and $L_2(s_k) = 0$. Thus, the zero spins ($s_k = 0$) are unstable. As each of them flips to $+1$, $L_1(s_k)\rightarrow L_1(s_k)+\frac{1}{N}$ and $-L_2(s_k)\rightarrow -L_2(s_k)+\frac{|K|}{N}$ and it continues as long as the condition: $-L_2(s_k) < L_1(s_k)$ holds. If that condition holds until all the zero spins flip to $+1$ (i.e) $\Delta<\frac{K(K+1)}{K-1}$, then the steady state $s_i=1$  $\forall i$ is reached at $H_c^{(1)}= \Delta/K$. If that condition stops holding before all the spins flip to $+1$, we reach an unsaturated steady state at $H_c^{(1)} = \Delta/K$ with $m = \frac{\Delta}{K}\left(\frac{K-1}{K+1}\right)$. After that, with $H$ increased by $\delta H$, a steady state is reached as each zero spin-flips to $+1$. The system eventually reaches the state $s_i=1$  $\forall i$ at $H_c^{(2)} = -K+\Delta-1$.\\ 
After reaching the state with $s_i=1$  $\forall i$,  $H$ is decreased quasi-statically.  Analogously, the backward coercive fields are $H_c^{(1)}= -\Delta/K$ and $H_c^{(2)} = K-\Delta+1$.

This results in $m-H$ plots with both first and second-order transitions. The hysteresis plot look like Fig. \ref{fig9} (e) and (f).
%  After reaching the state with $s_i=1$  $\forall i$,  the $H$ is decreased quasi-statically and now first spin  flips to $0$ at $H = -K+\Delta -1$ and all the spins will flip to $-1$  at $$H_c^{(2)} = -\Delta/K$$

\end{enumerate}
The Blume-Emery-Griffiths models Glauber steady state phase diagrams are very rich and depend non-trivially on the initial state. In Fig. \ref{fig10} we plot the phase diagram for $m=+1$ and $m=-1$ initial states to illustrate this difference and also the deviations from the equilibrium phase diagrams of Fig.  \ref{fig8}.

\subsubsection{with disorder}
\begin{figure*}[htbp]
\centering
\includegraphics[width=0.6\hsize]{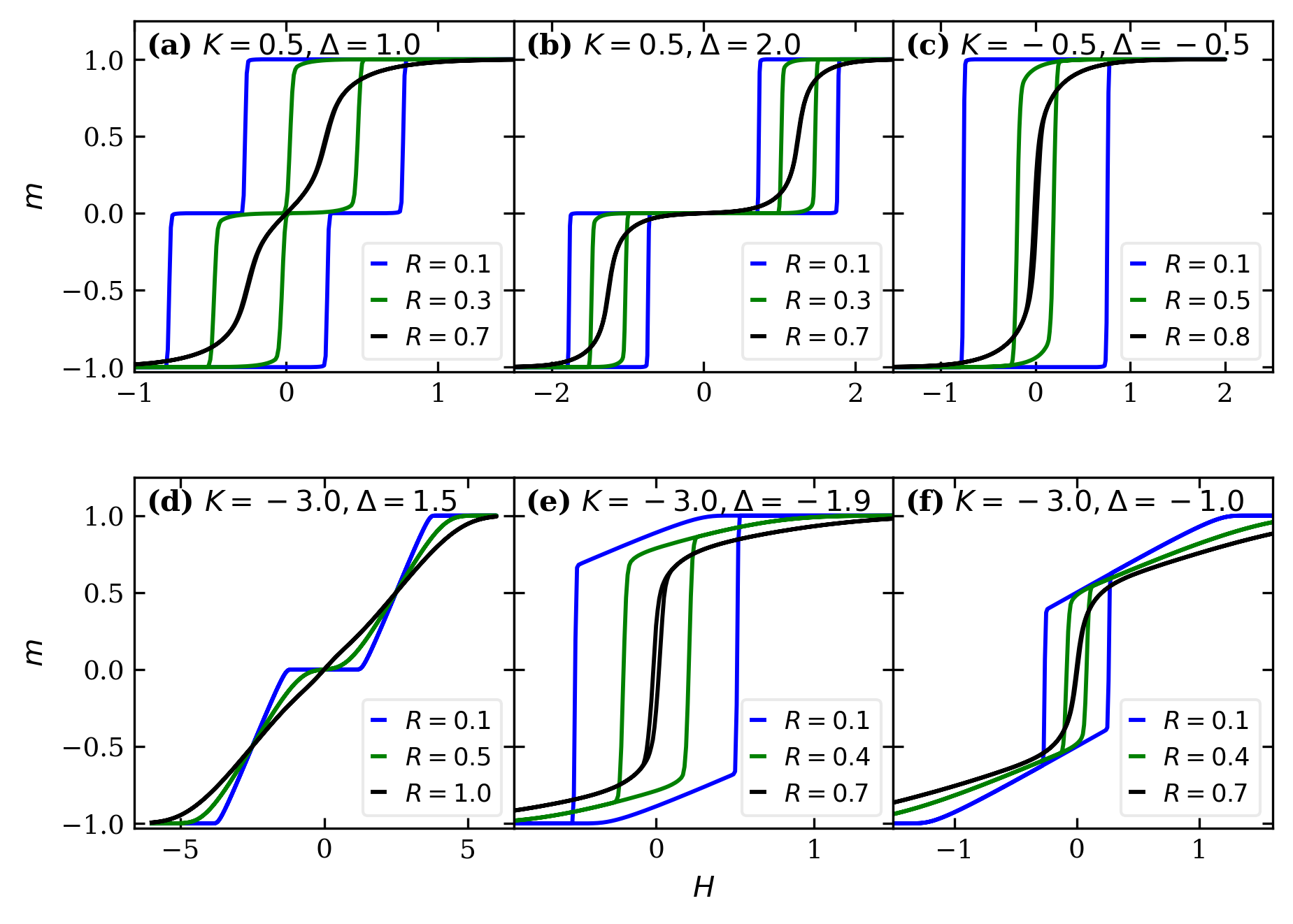}
\caption{The $m-H$ plots in the presence of disorder for RFBEGM for few representative values of $\Delta$ corresponding to different hysteresis loops in Fig. \ref{fig9}. The shape of the plot for $R=0$ is retained till the hysteresis vanishes a $R = R_c$. The data for all the plots are obtained for a system of size $N = 1000$, having averaged over $1000$ realizations of random field distributions for each value of $H$.}
\label{fig11}
\end{figure*}

In the case of RFBEGM the hysteresis loops shrink with $R$ till they vanish at a critical value of $R$. The critical value of the coercive field for a given $R$ can be obtained by solving the Eqs. 51 and 52 with  $m$ replaced by $m+H$, just like in the case of RFBCM. The shape  of the hysteresis loops are decided  by the  deterministic trajectories at $R=0$  under Glauber dynamics which were obtained in the previous sub-section. This is corroborated by numerically studying the $m-H$ hysteresis for different $R$. The resultant hysteresis loops are plotted in Fig. \ref{fig11}. Again the shape of the hysteresis loops is similar to Fig. \ref{fig9}, 
with hysteresis loop area decreasing with $R$ and  vanishing at  $R_c$. 

\section{Discussion}
\label{sec8}
We present a number of new results for the RFBCM and RFBEGM models in this paper, both for equilibrium and non-equilibrium Glauber steady states at zero temperature. We focused at $T=0$ as for finite $T$ both are essentially the same in the steady state. However, the dynamics at finite $T$ for the Glauber dynamics can still be  non-trivial due to the presence of random field. Our work can be extended to study the finite time dynamics using the results of the Sec. \ref{sec4}. For example, the time evolution of the magnetisation is given by the local slope of the function $I(m,q)$. 

The study in the absence of external field, revealed very rich phase diagrams and path dependent nature of the phase transitions in the two models. The effect of broken detailed balance though goes away at high enough random field disorder. This is similar to the effect of temperature on spin glasses \cite{derrida}. Hence we find that the variance of the disorder distribution can be used to tune the noise in disordered models. Even though the entire analysis in this paper is done for complete graphs, we expect similar phase diagram in three dimensions as, for pure Blume-Capel and Blume-Emery-Griffiths models the mean field captures the phase diagram for finite dimensions correctly.

We show that RFBEGM is a useful model to study the effect of frustration. Usually the frustration is studied via introducing  anti-ferromagnetic spin-spin coupling randomly like in spin glass models. In RFBEGM frustration is introduced via the repulsive biquadratic coupling. The RFBEGM comes out as an analytically tractable model to study the effect of frustration. The amount of frustration can be tuned by tuning $K$ in the model. Also the positive crystal field term introduces dilution by favoring $0$ spins. One novel interesting result is the crossover in the RFBEGM from a phase with no transition to a phase with order-disorder transition at the usual value of $R=R_c =\sqrt{\frac{2}{\pi}}$ in the region with $K<\Delta<0$ where there is a competition  as  the negative $K$ supports $0$ spins and crystal field suppresses $0$ spins. As a result $z=Kq-\Delta$ changes during the evolution and once $q$ is sufficiently small, $z$ changes sign and the crystal field dominates. The system exhibits dynamic crossover in its behavior from a frustrated system with no transition to the system in the non-equilibrium RFIM universality class due to the weakening repulsive interaction by the introduction of $0$ spins with decreasing $q$. Interestingly, this was also the regime where the model has non-abelian dynamics as shown in \cite{aldrin2}.
\begin{figure*}[htbp]
\centering
\includegraphics[width=0.5\hsize]{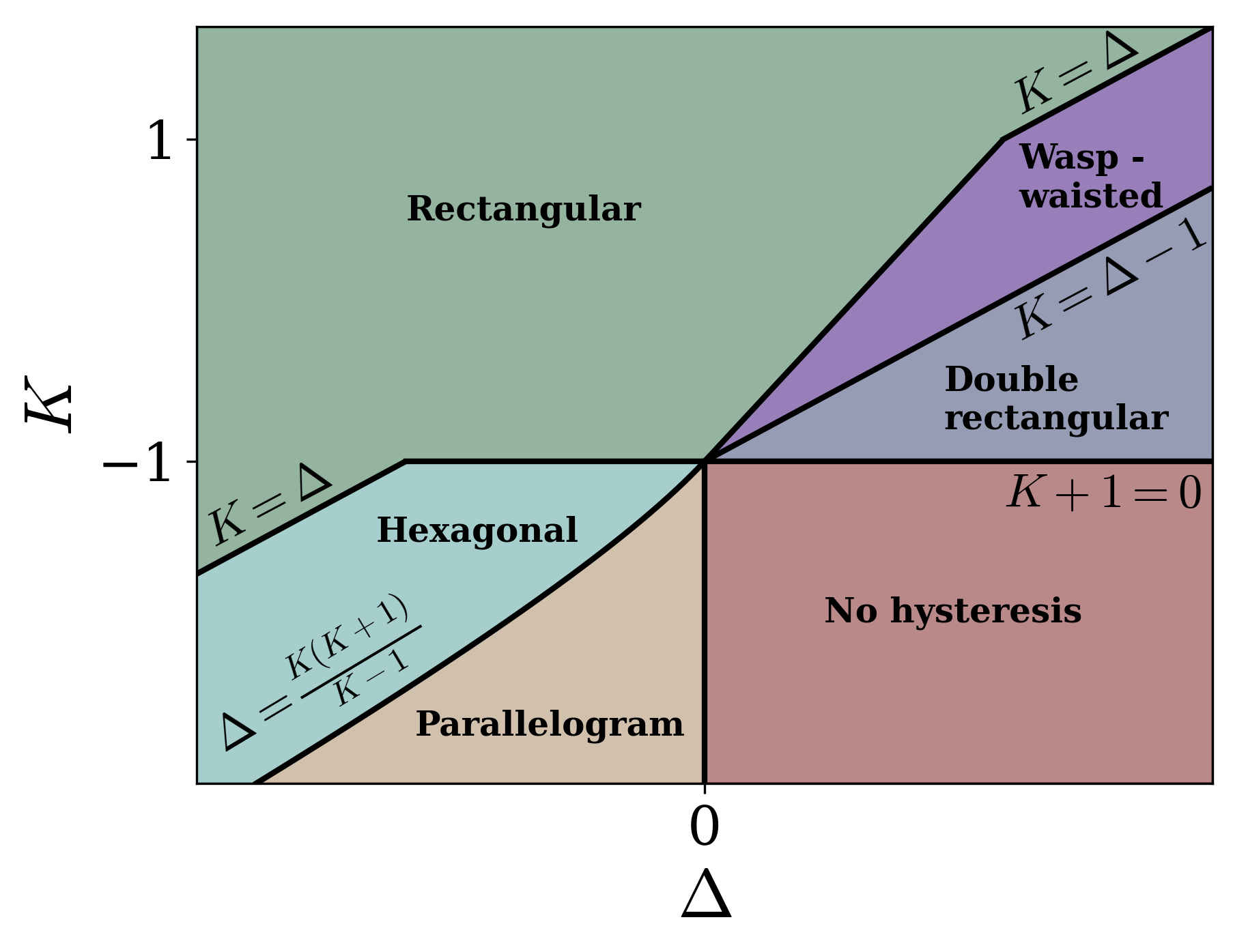}
\caption{Summary of different hysteresis loops observed in RFBEGM in $K-\Delta$ plane. The competing biquadratic coupling and crystal field results in loops that are parralllegram(also known as pot-bellied) or hexagonal in shape. This is also the region of dynamic crossover in the Glauber steady state as shown in Sec. \ref{sec6}. Similarly positive crystal field introduces non-magnetic sites and results into a wasp-waisted or double hysteresis loops.}
\label{fig12}
\end{figure*}

In the presence of external magnetic field we derived the shapes of hysteresis loops as a function of $K$ and $\Delta$ for RFBCM and RFBEGM. Some of these shapes have been reported earlier in simulations \cite{ortin1,wasp,aldrin2} and in experiments\cite{ferroelectricloops,ferro1,wasp1}. In this paper we obtain them analytically and connect the different hysteresis shapes with the underlying physics via our exact analysis. The results are summarized in Fig. \ref{fig12}. The wasp-waisted and pot-bellied hysteresis loops are  often reported in magnetic rocks and ferroelectric materials \cite{wasppot}. Our analysis connects these shapes to the presence of frustration and the presence of non-magnetic spins in the models.

{\bf{Data Availability}} The data supporting the findings of this study were generated using a C program together with instructions for reporducing the results and data, are available  from the authors upon reasonable request. 

{\bf{Conflicts of Interest}} The authors declare that they have no conflict of interest.

\end{document}